\documentclass[preprint, times]{elsarticle}
\journal{International Journal of Forecasting}

\usepackage[utf8]{inputenc}
\usepackage[a4paper]{geometry}
\usepackage{array}
\usepackage{booktabs}
\usepackage{bm}
\usepackage{amsthm}
\usepackage{amsmath}
\usepackage{amstext}
\usepackage{graphicx}
\usepackage{natbib}
\usepackage{hyperref}
\usepackage{lineno}

\usepackage{breakurl}

\hypersetup{
  colorlinks=true,
  linkcolor=bblue,
  citecolor=blue,
  urlcolor=blue
}

\usepackage{bm}

\usepackage{enumitem}
\begin{document}

\begin{frontmatter}

  \title{Improving forecasting performance using covariate-dependent copula models}

  \author[mainaddress]{Feng Li} \ead{feng.li@cufe.edu.cn}

  \author[secondaryaddress]{Yanfei Kang\corref{correspondingauthor}}
  \cortext[correspondingauthor]{Corresponding author} \ead{yanfeikang@buaa.edu.cn}

  \address[mainaddress]{School of Statistics and Mathematics, Central University of
    Finance and Economics, Beijing 100081, China.}

  \address[secondaryaddress]{School of Economics and Management, Beihang University,
    Beijing 100191, China.}

  \begin{abstract}
    Copulas provide an attractive approach for constructing multivariate distributions
    with flexible marginal distributions and different forms of dependences. Of particular
    importance in many areas is the possibility of explicitly forecasting the
    tail-dependences. Most of the available approaches are only able to estimate
    tail-dependences and correlations via nuisance parameters, but can neither be used for
    interpretation, nor for forecasting. Aiming to improve copula forecasting performance,
    we propose a general Bayesian approach for modeling and forecasting tail-dependences
    and correlations as explicit functions of covariates. The proposed covariate-dependent
    copula model also allows for Bayesian variable selection among covariates from the
    marginal models as well as the copula density. The copulas we study include
    Joe-Clayton copula, Clayton copula, Gumbel copula and Student's
    \emph{t}-copula. Posterior inference is carried out using an efficient MCMC simulation
    method. Our approach is applied to both simulated data and the S\&P 100 and S\&P 600
    stock indices.  The forecasting performance of the proposed approach is compared with
    other modeling strategies based on log predictive scores.  Value-at-Risk evaluation is
    also preformed for model comparisons.

  \end{abstract}


\end{frontmatter}


\setcounter{page}{1}

\section{Introduction}
\label{sec:intro}

Copula modeling is an active research area dating back to Sklar's Theorem
\citep{sklar1959fonctions}. The theorem states that any multivariate cumulative
distribution function can be written in terms of univariate marginal distributions and a
copula function. Allowing the researchers to specify the marginal distributions separately
from the copula function that links theses distributions to form the joint distribution,
copula models provide a great deal of flexibility in modeling multivariate distributions,
and thus become very popular in multivariate analysis. The related studies include, but
are not limited to, an introduction to copulas \citep{nelsen2006introduction}, research on
dependences and extreme value distributions with copulas \citep{joe1997multivariate},
research on the construction of multivariate dependences using bivariate copulas
\citep{kurowicka2010dependence}, and approaches in copula estimation (see
\citet{jaworski2010copula} for a review) including pair-copula constructions (see
\citet{czado2010pair} and references therein).

Copula models have been widely used in forecasting as well as in risk evaluation due to
their capacity to model tail-dependences and correlations (see \citet{patton2012copula}
and \citet{smith2015copula} for recent surveys).  \citet{dobric2005nonparametric} use
nonparametric methods to estimate the lower tail-dependence in bivariate copulas with
continuous marginals for assets in German DAX30 stock index. \citet{schmidt2006non}
explore the tail-dependence estimators for the tail copula with exchange rates for
Deutsche Mark to US dollars, and Japanese Yen to US dollars.  \citet{draisma2004bivariate}
use hypothesis testing to detect the dependence of extreme events when they are
asymptotically independent. A case study for rain-wind data of modeling the asymptotic
dependence can be found in \citet{ledford1997modelling}. Tail dependences result in the
tail risk of value-at-risk (VaR).  One of the main difficulties in forecasting VaR of
portfolio is to model the tail-dependence structure and forecast the future
tail-dependence. Extensive empirical studies (e.g., \citet{huang2009estimating},
\citet{siburg2015forecasting}) show that a copula model with accurately specified
tail-dependence structure, especially in the lower tail-dependence, yields better VaR
forecasting.

In copula-based tail-dependence forecasting, letting the tail-dependences and the
Kendall's $\tau$ in copula modeling be fixed numbers is very restrictive. This is
particularly true in financial time-series applications, where the tail-dependences have
been shown to vary with time \citep[e.g.,][]{patton2006modelling} and be affected by
volatility information \citep[e.g.,][]{poon2003extreme}. Therefore, it is natural to
condition on the available information, and thus we are led to study the conditional
structure in tail-dependences.  Despite the vast literature in tail-dependence estimation,
little attempts have been made to investigate the factors that affect the variation of
tail-dependences, not to mention using these factors to model tail-dependences and improve
forecasting performance. This is partially because the computational complexity is still a
challenge in many situations. Moreover, most approaches for modeling rank correlations and
tail-dependences using existing copulas require firstly modeling intermediate parameters
and thereafter obtaining them, which is inconvenient to link the correlations and
tail-dependences directly with available covariates.

In this paper, we use \emph{feature} to describe a characteristic in the univariate and
multivariate densities (e.g. Kendall's $\tau$ and tail-dependence are two features of a
copula, while mean, variance, skewness and kurtosis are features of a univariate density).
We propose a general Bayesian approach for copula modeling and forecasting with
explanatory variables introduced in both tail-dependences and Kendall's $\tau$ of the
copula function. This construction allows us to explore the drivers of different forms
of dependences, and obtain more informative forecasting based on
available information. Variable selection is carried out in the features of marginal models as
well as the copula features. We propose an efficient Markov chain Monte Carlo (MCMC)
simulation method for posterior inference which is also used to calculate the predictive
densities based on posterior draws.

The outline of the paper is as follows. Section \ref{sec:copula-model} introduces the
covariate-dependent copula model. We discuss the prior specification for the model and
present the general form of the inferential approach in Section \ref{sec:prior}.  Section
\ref{sec:mcmc} presents the details of the MCMC scheme. In Section \ref{sec:simulation},
we demonstrate the advantages of including covariates in the dependences by evaluating the
out-of-sample forecasting performances on simulated data. In Section
\ref{sec:application}, we use our proposed model to forecast the daily returns on the S\&P
100 and S\&P 600 stock market indices and make comparisons with other competing modeling
strategies. Section \ref{sec:conclusion} concludes the paper and discusses some potential
directions for further research. The appendix documents the computational details used in
the MCMC scheme.

\section{The covariate-dependent copula model}
\label{sec:copula-model}

A copula is a multivariate distribution that separates the univariate marginal
distributions and the multivariate dependence structure. The correspondence between a
multivariate distribution $F(x_{1},...,x_{M})$ and a copula function $C(u_{1},...,u_{M})$
can be expressed as
\begin{align*}
  F(x_{1},...,x_{M})=F(F_{1}^{-1}(u_{1}),...,F_{M}^{-1}(u_{M}))=C(u_{1},...,u_{M}
  )=C(F_{1}(x_{1}),...,F_{M}(x_{M})),
\end{align*}
where the correspondence is one-to-one with continuous marginal distributions.  Copulas
provide a general approach for constructing flexible multivariate densities
\citep{joe1997multivariate}. For example, the bivariate Gaussian copula
$C(u_{1},u_{2}|\rho)=\Psi(F_{1}^{-1}(u_{1}),F_{2}^{-1}(u_{2})|\rho)$, with the correlation
$\rho$, is a relaxed Gaussian density in the sense that $F_{1}(\cdot)$ and $F_{2}(\cdot)$
do not need to be normal \citep{pitt2006efficient}.

\subsection{Dependence concepts}

Modeling the multivariate dependence typically involves quantifying two important
quantities: tail-dependence and correlation. A key feature in copula models is that the
multivariate dependence does not depend on marginal densities.

In copula models, the tail-dependence describes the concordance between extreme values of
random variables $X_{1}$ and $X_{2}$. The lower tail-dependence $\lambda_{L}$ and the
upper tail-dependence $\lambda_{U}$ can be expressed in terms of bivariate copulas:
\begin{align*}
  \lambda_{L}=&\lim_{u\to0^{+}}p(X_{1}<F_{1}^{-1}(u)|X_{2}<F_{2}^{-1}(u))=\lim_
                {u\to0^{+}}\frac{C(u,u)}{u},\\
  \lambda_{U}=&\lim_{u\to1^{-}}p(X_{1}>F_{1}^{-1}(u)|X_{2}>F_{2}^{-1}(u))=\lim_
                {u\to1^{-}}\frac{1-2u+C(u,u)}{1-u}.
\end{align*}
Not all copulas have both lower and upper tail-dependences. For single-parameter copulas,
which is only able to capture one side of the tail dependences, like the Clayton and the
Gumbel copula, a simple way to achieve the same effect is to rotate the copula by 180
degrees.

The correlation between two variables is usually measured with rank correlations such as
Kendall's $\tau$:
\begin{align*}
  \tau=4\int\int F(x_{1},x_{2})\mathrm{d}F(x_{1},x_{2})-1=4\int\int
  C(u_{1},u_{2})\mathrm{d}C(u_{1},u_{2})-1.
\end{align*}
An attractive property of Kendall's $\tau$ is that it is invariant with respect to
strictly monotonic transforms. In this paper, we focus on modeling the correlation in
terms of Kendall's $\tau$. Other types of correlations, like Spearman's rank correlation,
can be equally well modeled with our method. For measuring the dependences in trivariate
distributions, one may consider using a three-dimensional version of correlations
\citep{garcia2013new}. Correlations in higher dimensions are usually estimated pairwisely.

\subsection{The reparameterized copula}
\label{sec:bb7}

In this section, we describe how to reparameterize the copula function so that the
parameters highlight the copula features of interest. The default parameters in most
copula functions do not directly represent the copula features, such as the
tail-dependences and Kendall's $\tau$. We demonstrate the parameterization with the widely
used two-parameter Joe-Clayton copula, which is flexible to be reparameterized in terms of
any pairwise combination of lower tail-dependence, upper tail-dependece, and Kendal's
$\tau$.

Our copula modeling approach is general and can be applied to model any copulas. In
Section \ref{sec:application}, as well as our computer program, we extend our model
class to Clayton copula, Gumbel copula, and Student's \emph{t}-copula. The popular vine
copula construction \citep{aas2009pair,czado2012maximum} can also be used to extend our
bivariate copula modeling to higher dimensions with discrete and continuous margins, but
requires further extensive work and will be discussed in a separate paper.

\subsubsection{The Joe-Clayton copula}
\label{sec:bb7density}

The Joe-Clayton copula function (also known as the BB7 copula), introduced by
\citet{joe1997multivariate}, can be written as
\begin{align*}
  C(u,v|\theta,\delta)= & \eta(\eta^{-1}(u)+\eta^{-1}(v))\\
  = & 1-\left[1-\left\{
      \left(1-\bar{u}^{\theta}\right)^{-\delta}+\left(1-\bar{v}^{\theta}\right)^{
      -\delta}-1\right\} ^{-1/\delta}\right]^{1/\theta},
\end{align*}
where $\eta(s)=1-[1-(1+s)^{-1/\delta}]^{1/\theta}$, $\theta\geq1$, $\delta>0$,
$\bar{u}=1-u$, and $\bar{v}=1-v$.  The copula density function for the Joe-Clayton copula
is
\begin{equation}
  \begin{split}
    \label{eq:bb7-density}
    c(u,v|\theta,\delta)=\frac{\partial^{2}C(u,v,\theta,\delta)}{\partial u\partial v}= &
    \left[T_{1}(u)T_{1}(v)\right]^{-1-\delta}T_{2}(u)T_{2}(v)L_{1}^{-2(1+\delta)
      /\delta}\\
    & \times(1-L_{1}^{-1/\delta})^{1/\theta-2}\left[(1+\delta)\theta
      L_{1}^{1/\delta}-\theta\delta-1\right],
  \end{split}
\end{equation}
where $T_{1}(s)=1-(1-s)^{\theta}$, $T_{2}(s)=(1-s)^{\theta-1}$ and
$L_{1}=T_{1}(v)^{-\delta}+T_{1}(u)^{-\delta}-1$.

The lower and upper tail-dependences for Joe-Clayton copula are
$\lambda_{L}=2^{-1/\delta}$ and $\lambda_{U}=2-2^{1/\theta}$, respectively.  Previous work
for Joe-Clayton copula includes the time-varying dependence modeling with symmetric
version of the Joe-Clayton copula \citep{patton2006modelling} and the development of
multivariate dependence modeling via a vine structure based on bivariate copulas such as
the Joe-Clayton copula \citep[e.g.,][]{aas2009pair, czado2012maximum}.

\begin{figure}
  \centering \includegraphics[width=0.7\textwidth]{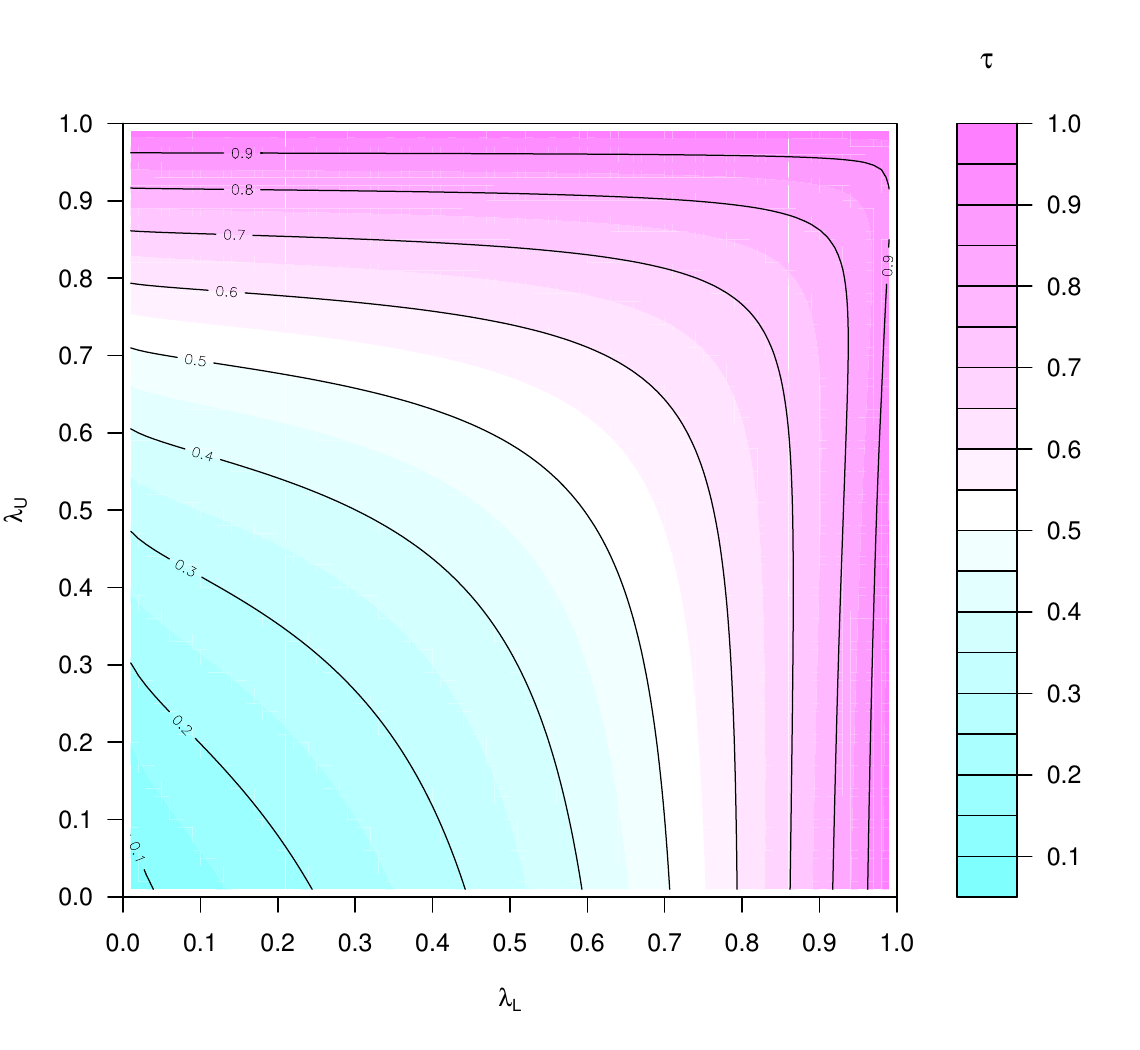}
  \caption{The contour plot of the Kendall's $\tau$ with respect to lower tail-dependence
    ($\lambda_{L}$) and upper tail-dependence ($\lambda_{U}$) for the Joe-Clayton copula.}
  \label{fig:tau}
\end{figure}

The Kendall's $\tau$ of the Joe-Clayton copula for the case $1\leq\theta<2$ can be found
in \citet{smith2012estimation}, where only moderate level of correlation is possible. We
extend it to $\theta\geq 2$ and present the full expression for all $\theta\geq1$:
\begin{align*}
  \tau(\theta,\delta)=\begin{cases}
    1-2/[\delta(2-\theta)]+4B\left(\delta+2,2/\theta-1\right)/(\theta^{2}\delta) &
    1\leq\theta<2;\\
    1-\left[\psi(2+\delta)-\psi(1)-1\right]/\delta & \theta=2;\\
    1-2/[\delta(2-\theta)]-4\pi/\left[\theta^{2}\delta(2+\delta)\sin(2\pi/\theta
      )B\left(1+\delta+2/\theta,2-2/\theta\right)\right] & \theta>2,
  \end{cases}
\end{align*}
where $B(\cdot)$ is the beta function and $\psi(\cdot)$ is the digamma function.  By the
dominated convergence theorem, Kendall's $\tau$ is continuous for all $\theta\geq1$. If we
employ the equations $\delta=-\log2/\log\lambda_L$ and $\theta=\log2/\log(2-\lambda_{U})$
from the previous results, we can rewrite Kendall's $\tau$ in terms of lower and upper
tail-dependences as $\tau(\lambda_L,\lambda_{U})$. See Figure~\ref{fig:tau} for a contour
plot of these relationships.

\subsubsection{Reparametrization}
\label{sec:bb7-rep}

To simplify the interpretation of the features in a copula model, we parameterize it in
terms of the lower tail-dependence $\lambda_{L}$ and Kendall's $\tau$,
\begin{align*}
  C(u,v|\lambda_{L},\tau)= & 1-\left[1-\left\{
                             \left[1-\bar{u}^{\log2/{\log(2-\tau^{-1}(\lambda_{L}))}}\right]^{\log2/
                             {\log\lambda_{L}}}\right.\right.\\
                           &
                             \hspace{1.8cm}\left.\left.+\left[1-\bar{v}^{\log2/{\log(2-\tau^{-1}(\lambda_{L}
                             ))}}\right]^{\log2/{\log\lambda_{L}}}-1\right\}
                             ^{\log\lambda_{L}/\log2}\right]^{{\log(2-\tau^{-1}(\lambda_{L}))}/\log2},
\end{align*}
where $\tau^{-1}(\lambda_{L})=\lambda_{U}=2-2^{1/\theta}$ is the inverse function of
Kendall's $\tau$ given $\lambda_{L}$.  The related reparametrized copula density is
obtained by substituting $\delta=-\log2/\log\lambda_{L}$ and
$\theta=\log2/\log(2-\tau^{-1}(\lambda_{L}))$ in Equation (\ref{eq:bb7-density}).

There are also alternative reparametrization schemes that allow for modeling lower and
upper tail dependences $(\lambda_L,\lambda_U)$, or upper tail dependence and Kendall's
$\tau$ $(\lambda_U,\tau)$, as shown in our simulation study and real data application.  In
principle, there is no difference among using parameterization in terms of
$(\lambda_L,\tau)$, $(\lambda_L,\lambda_U)$ and $(\lambda_U,\tau)$ with regard to the
Joe-Clayton copula if we are only interested in modeling the features as constants. When any
two of the three parameterizations are used, the third can be easily derived due to the
deterministic relationships among lower tail-dependence, upper tail-dependence and
Kendall's $\tau$, as shown in Figure~\ref{fig:tau}.

Our parameterization has two primary advantages. Firstly, it makes it easier to specify
the prior information in the Bayesian approach highlighted in Section~\ref{sec:prior}.
Secondly, and more importantly, the parameterization makes it possible to directly link
correlations and tail-dependences to covariates, which yields conditional forecasting of
the dependences. See Section~\ref{sec:covariate-dep-copula} for the details.

\subsection{Covariate-dependent copula features}
\label{sec:covariate-dep-copula}

We now introduce a covariate-dependent copula model that allows the copula features to be
linked to the observed covariates.  A prominent example is the covariate-dependent correlation
and tail-dependences in a bivariate copula:
\begin{align}
  \label{eq:tail-dep}
  \tau=l_{\tau}^{-1}(\bm{X}\bm{\beta}_{\tau}),~\mathrm{and}~\lambda=l_{\lambda}^{
  -1}(\bm{X}\bm{\beta}_{\lambda}),
\end{align}
where $\lambda$ without subscripts represents the dependences in the lower and/or upper
tails; $\tau$ is Kendall's $\tau$; $\bm{X}$ is the set of covariates matrix used in the
two margins and $\bm{\beta}$ with subscripts is the corresponding coefficients
vector. Furthermore, $l_{\tau}(\cdot)$ and $l_{\lambda}(\cdot)$ are suitable link
functions that connect $\lambda$ and $\tau$ with $\bm{X}$.

This covariate-dependent parameterization has three primary advantages. First, our
approach makes it possible to use all marginal information to model the tail-dependences,
which also yields the conditional forecasting on tail-dependences.  Second, this
parameterization can be applied to a rich class of copulas. Besides, ARMA-like
variation in tail-dependences \citep{patton2006modelling} and GARCH-like dependences in
the dependences \citep{lucas2014conditional} can be considered as special cases of our
model since their lagged dependences and volatility variables can be used as the
covariates in our model. Third, it is straightforward to construct the likelihood function
in our model (See Section \ref{sec:mcmc} for the MCMC details) and variable selection can
be used to select meaningful covariates that influence the dependences and also prevent
overfitting.

Compared with the bivariate DCC-GARCH model \citep{engle2002dynamic} and the bivariate
volatility model \citep{yu2006multivariate}, our approach not only leads to a more
insightful interpretation of the features, but allows for conditional heteroscedasticity
in the dependences and generates more accurate forecasts as a result. See a detailed simulation
study in Section \ref{sub:Model-comparison-dgp} and real data forecasting analysis in
Section \ref{sub:Model-comparison-sp}.

\subsection{Marginal models}
\label{sub:marginal-models}

In this paper we use\emph{ margins} as a synonym for marginal models.  In principle, the
copula approach can be used with any margins, but we assume the margins to be
split-\emph{t} distributions \citep{li2010flexible} in our simulation studies in Section
\ref{sec:simulation} and the real data application in Section \ref{sec:application}. The
split-\emph{t} is a flexible four-parameter distribution with the Student's \emph{t}
distribution, the asymmetric normal and the symmetric normal distributions as its special
cases. See \citet{li2010flexible} for its properties.

Following \citet{li2010flexible}, we allow the mean $\bm{\mu}_k$, the scale $\bm{\phi}_k$,
the degrees of freedom $\bm{\nu}_k$ and the skewness $\bm{\kappa}_k$ of the split-$t$
density in the \emph{k}:th margins to be linked to covariates in the following matrix
form:
\begin{align}
  \label{eq:margin-dep}
  \bm{\mu}_{k}=\bm{X}_{k}\bm{\beta}_{\mu_{k}},~\bm{\nu}_{k}=\exp(\bm{X}_{k}\bm
  {\beta}_{\nu_{k}}),
  ~\bm{\phi}_{k}=\exp(\bm{X}_{k}\bm{\beta}_{\phi_{k}}),~\mathrm{and}~\bm{\kappa}_
  {k}=\exp(\bm{X}_{k}\bm{\beta}_{\lambda_{k}}),
\end{align}
where $\bm{X}_{k}$ is the covariate matrix in the \emph{k}:th margin.

One may also consider using mixture models in the margins.  \citet{li2010flexible} shows
in an application to S\&P 500 data that the one-component split-$t$ model with all
features linked to covariates does well in comparison with mixtures of split-$t$
components.  In this paper, we use the one-component split-\emph{t} model
for demonstration purposes. Our inference based on covariate-dependent structures is
straightforward and also eases interpretation. Furthermore, compared with other volatility
margins, the split\emph{-t} margin can explicitly model the volatility, kurtosis and
skewness with linked covariates.

Our inference procedure can be generally applied to different margins. We also implement
the univariate GARCH model \citep{bollerslev1986generalized} and the univariate stochastic
volatility (SV) model \citep{melino1990pricing} as margins for the purpose of comparison
in Section~\ref{sub:Model-comparison-sp}. The GARCH model and the SV model are popularly
used in modeling volatility. The major difference between the two approaches is that
volatility is conditionally deterministic in the GARCH model but is modeled via a latent
process in the SV model. It is worth mentioning that implementing the univariate and
multivariate GARCH-type models requires carefully calibrating their likelihood functions
and constraints. Moreover, the copula modeling requires the calculation of CDF from
marginal densities. However, this calculation is not always analytical. The Bayesian
approach allows us to calculate the CDF for the GARCH and SV models with Monte Carlo
integration when no analytical CDF applies.

\section{The priors and posterior}
\label{sec:prior}

We use the same technique to specify priors for marginal and copula features.  We omit the
subscripts that indicate the functionality of the parameters in this section for
convenience. With our parameterization, the parameters to be estimated are the
coefficients $\bm{\beta}$ in Equations (\ref{eq:tail-dep})-(\ref{eq:margin-dep}) and the
variable selection indicator $\bm{\mathcal{I}} = [\mathcal{I}_1,...,\mathcal{I}_j,...]$
defined below.

Letting $\mathcal{I}_j$ be the variable selection indicator for a given covariate, we
express it as
\begin{align*}
  \mathcal{I}_{j}=\begin{cases}
    1, & \text{if the $j$:th variable is included in  the model},\\
    0, & \text{otherwise}.
  \end{cases}
\end{align*}
We standardize each covariate to have zero mean and unit variance. The prior for each
variable selection indicator is identically distributed as $Bern(p)$ where $p$ is the
hyperparameter. We assume that the variable selection indicator for the intercept is one.
Note that when the number of covariates is large, the independent Bernoulli prior can be
very informative \citep{yau2003bayesian}. We also implement the recommended $Beta$ prior
\citep{scott2010bayes} where the uniform prior is a special case. Our comparison in
simulation studies show that the difference between independent Bernoulli prior and
uniform prior is negligible with less than twenty covariates.

We partition the coefficients vector $\bm{\beta}$ into the intercept $\beta_{0}$ and the
slopes vector $\bm{\beta}_s$, which have independent priors. We then decompose the joint prior
as
\begin{align*}
  p(\bm{\beta},\bm{\mathcal{I}})=
  p(\beta_{0},\bm{\beta}_s,\bm{\mathcal{I}})=p(\beta_{0})p(\bm{\beta}_s,\bm{\mathcal{I}})=p(\beta_{0}
  )p(\bm{\beta}_s|\bm{\mathcal{I}})p(\bm{\mathcal{I}}).
\end{align*}
We use normal priors for both $\beta_{0}$ and $\bm{\beta}_s$ conditional on variable
selection indicators $\bm{\mathcal{I}}$. The mean and variance (covariance matrix) in normal
priors for $\beta_{0}$ and $\bm{\beta}_s$ are specified in Section \ref{sec:prior-int} and
Section \ref{sec:prior-slop-vs}, respectively.

\subsection{Prior mean and variance for $\beta_0$}
\label{sec:prior-int}

We set the prior for the intercept by following the strategy in
\citet{villani2012generalized}. We use the prior belief on the copula and marginal
features to derive the implied prior on the intercept $\beta_{0}$ under the assumption
that the covariates effect is zero. This technique can be applied to the following three
types of link functions directly.

\begin{enumerate}[noitemsep]

\item When the identity link is used, setting the implied prior on the model parameter is
  trivially the same as on the intercept.

\item When the link is the log function, assuming a log-normal distribution on the model
  parameter with hyperparameters mean $m$ and variance $\sigma^{2}$  yields a normal
  prior with mean $\log(m)-\log[\sigma^{2}/m^{2}+1]/2$ and variance
  $\log[\sigma^{2}/m^{2}+1]$ in the intercept.

\item When the logit link is used, we take the tail-dependence $\lambda$ as an example.
  When there is only an intercept in the covariates, we have
  $\lambda=1/(1+\exp(-\beta_{0}))$. Therefore, if we assume $\lambda$ follows the beta
  distribution $Beta(m,\sigma^{2})$ with its hyperparameters mean $m$ and variance
  $\sigma^{2}$, we have the mean and variance of $\beta_{0}$ as:
  \begin{align*}
    E(\beta_{0}) &
                   =\int_{0}^{1}\log(\frac{x}{1-x})Beta(x,m,\sigma^{2})\mathrm{d}x=\psi(\alpha_{1}
                   )-\psi(\alpha_{2}),\\
    Var(\beta_{0}) &
                     =\int_{0}^{1}(\log(\frac{x}{1-x}))^{2}Beta(x,m,\sigma^{2})\mathrm{d}x-E^{2}
                     (\beta_{0})=\psi_{1}(\alpha_{1})+\psi_{1}(\alpha_{2}),
  \end{align*}
  where $\psi(\cdot)$ and $\psi_{1}(\cdot)$ are the digamma and trigamma functions, $\alpha_{1}=-m(m^{2}-m+\sigma^{2})/\sigma^{2}$ and
  $\alpha_{2}=-1+m+(m-1)^{2}m/\sigma^{2}$. We can now set the prior on the intercept
  $\beta_{0}$ based on hyperparameters.

\end{enumerate}

\subsection{Prior mean and covariance matrix for $\beta_s$ given $\mathcal{I}$}
\label{sec:prior-slop-vs}

We first consider the case without variable selection. We assume that the slopes $\bm{\beta}_s$
are normally distributed as $N(0, \bm{\Sigma})$ where $\bm{\Sigma}=c^{2}\cdot \bm{P}^{-1}$
is the covariance matrix, $\bm{P}$ is a positive definite symmetric matrix, and $c$ is a
scaling hyperparameter. In the application, $\bm{P}$ is the identity matrix.  Using the
identity matrix can be sensitive to the scaling of covariates. This is alleviated by
standardizing the covariates in our application.

We then consider the case with variable selection.  Conditional on the variable selection
indicators, the slopes are still normally distributed with mean
$\bm{\mu}_{\mathcal{I}}+\bm{\Sigma}_{21}\bm{\Sigma}_{\mathcal{I}^{c}}^{-1}(\bm
{\beta}_{\mathcal{I}^{c}}-\bm{\mu}_{\mathcal{I}^{c}})$ and the covariance matrix becomes
$\bm{\Sigma}_{\mathcal{I}}-\bm{\Sigma}_{21}\bm{\Sigma}_{\mathcal{I}^{c}}^{-1
}\bm{\Sigma}_{12}$ \citep{mardia1979multivariate}, with obvious notations.

\subsection{The joint posterior}

The posterior in the copula model can be written in terms of the likelihoods from the
marginal distributions, the copula likelihood, and the prior for parameters in the copula
and marginal distributions as
\begin{align*}
  \log p(\{\bm{\beta},\bm{\mathcal{I}}\}|\bm{y},\bm{x})=
  &~ \mathrm{constant}+\sum\nolimits _{j=1}^{M}\log
    p(\bm{y}_{.j}|\{\bm{\beta},\bm{\mathcal{I}}\}_{j},\bm{x}_{j})\\
  &
    +\log\mathcal{L}_{C}(\bm{u}|\{\bm{\beta},\bm{\mathcal{I}}\}_{C},\bm{y},\bm{x})
    +\log p(\{\bm{\beta},\bm{\mathcal{I}}\}),
\end{align*}
where $\log p(\bm{y}_{.j}|\{\bm{\beta},\bm{\mathcal{I}}\}_{j},\bm{x}_{j})$ is the log
likelihood in the $j$:th margin, and the sets $\{\bm{\beta},\bm{\mathcal{I}}\}_{j}$ are
the parameter blocks in the $j$:th margin. Furthermore,
$\bm{u}=\left(\bm{u}_{1},...,\bm{u}_{M}\right)$, where
$\bm{u}_{j}=\left(u_{1j},...,u_{nj}\right)'$, $u_{ij}=F_{j}(y_{ij}|\{\beta,I\}_j)$.
$F_{j}(\cdot)$ is the CDF of the $j$:th marginal distribution, and $\mathcal{L}_{C}$ is
the likelihood for the copula function.  In our application, we have $M=2$ and we use the
reparametrized Joe-Clayton copula, Clayton copula, Gumbel copula, and Student's
\emph{t}-copula.

\section{Posterior inference and forecasting evaluation}
\label{sec:mcmc}

\subsection{Posterior inference}

The joint posterior is not tractable and we use the Metropolis--Hastings algorithm within
a Gibbs sampler. We update the copula component together with the marginal components
jointly. The Gibbs sampler is used for updating the joint parameter components, with each
conditional parameter block $\{\bm{\beta},\bm{\mathcal{I}}\}$ updated by the
Metropolis--Hastings algorithm. The complete updating scheme is as follows.

Following the updating order  in Table~\ref{tab:gibbs}, we jointly
update the coefficients and variable selection indicators
$\{\bm{\beta},\bm{\mathcal{I}}\}$ in each parameter block using an efficiency tailored
Metropolis--Hastings algorithm with integrated finite-step Newton proposals. The
acceptance probability for a proposed draw, $\{\bm{\beta}^{(p)},\bm{\mathcal{I}}^{(p)}\}$,
conditional on the current value of the parameters
$\{\bm{\beta}^{(c)},\bm{\mathcal{I}}^{(c)}\}$ is
\begin{align*}
  \min\left[1,\frac{p(\{\bm{\beta}^{(p)},\bm{\mathcal{I}}^{(p)}\}|\{\bm{\beta}^{
  (c)},\bm{\mathcal{I}}^{(c)}\},\bm{y},\bm{x})g(\{\bm{\beta}^{(c)},\bm{\mathcal{I
  }}^{(c)}\}|\{\bm{\beta}^{(p)},\bm{\mathcal{I}}^{(p)}\})}{p(\{\bm{\beta}^{(c)}
  ,\bm{\mathcal{I}}^{(c)}\}|\{\bm{\beta}^{(p)},\bm{\mathcal{I}}^{(p)}\},\bm{y}
  ,\bm{x})g(\{\bm{\beta}^{(p)},\bm{\mathcal{I}}^{(p)}\}|\{\bm{\beta}^{(c)},\bm
  {\mathcal{I}}^{(c)}\})}\right],
\end{align*}
where
$g(\{\bm{\beta}^{(p)},\bm{\mathcal{I}}^{(p)}\}|\{\bm{\beta}^{(c)},\bm{\mathcal
  {I}}^{(c)}\}$ is the proposal distribution in the Metropolis-Hastings algorithm for
$\{\bm{\beta},\bm{\mathcal{I}}\}$ conditional on the current draw. It can be decomposed as
$g(\{\bm{\beta}^{(p)},\bm{\mathcal{I}}^{(p)}\}|\{\bm{\beta}^{(c)},\bm{\mathcal
  {I}}^{(c)}\}=g_{1}(\bm{\beta}^{(p)}|\{\bm{\beta}^{(c)},\bm{\mathcal{I}}^{(p)}\}
_{i})\times g_{2}(\bm{\mathcal{I}}^{(p)}|\{\bm{\beta}^{(c)},\bm{\mathcal{I}}^{(c)}\})$.
Note that, $g_{1}(\bm{\beta}^{(p)}|\{\bm{\beta}^{(c)},\bm{\mathcal{I}}^{(p)}\}_{i})$ is
the proposal distribution where the proposal mode is from a finite-step Newton
approximation of the posterior distribution (usually smaller than three steps) starting on
the current draw, and the proposal covariance matrix is from the negative inverse Hessian
matrix.  In our application,
$g_{1}(\bm{\beta}^{(p)}|\{\bm{\beta}^{(c)},\bm{\mathcal{I}}^{(p)}\}_{i})$ is a
multivariate Student's \emph{t} distribution with six degrees of freedom.  The notion
$g_{2}(\bm{\mathcal{I}}^{(p)}|\{\bm{\beta}^{(c)},\bm{\mathcal{I}}^{(c)}\})$ is the
proposal distribution for variable's indicator vector $\bm{\mathcal{I}}$. We propose a
change of the $j$:th variable selection indicator $\mathcal{I}_j$ with probability
$p_{prop}=0.5$ in each MCMC iteration, i.e.,
$p_{prop}(\mathcal{I}_j^{(p)}=1 | \mathcal{I}_j^{(c)}=0)=p_{prop}(\mathcal{I}_j^{(p)}=0 |
\mathcal{I}_j^{(c)}=1)=0.5$. Our simple scheme works well in the copula model. See
\citet{nott2005adaptive} for alternative types of variable selection schemes, and see
\citet{tran2014copula}) for variational inference.

The finite-step Newton proposal was originally used in
\citet{villani2009regression,villani2012generalized} for univariate mixture models where
the gating function and parameters in mixing components are parameterized in terms of
covariates. It is shown in their applications that the Metropolis--Hastings algorithm with
finite-step Newton proposals increases the convergence rate rapidly. We extend the
algorithm to our covariate-dependent copula models which only requires the gradient for
the marginal distribution and copula model with respect to their features.
\ref{sec:chain-rule} documents the details for calculating the gradient with respect to
the copula features for reparametrized copulas in the MCMC implementation.

\begin{table}
  \begin{center}
    \caption{The Gibbs sampler for covariate-dependent copula. The notation
      $\{\bm{\beta}_{\mu},\bm{\mathcal{I}}_{\mu}\}_{m}$ denotes the covariate coefficients
      and variable selection indicators in copula component $m$ for feature $\mu$. The
      notation $\{\bm{\beta}_{\mu},\bm{\mathcal{I}}_{\mu}\}_{-m}$ indicates all the other
      parameters in the model except
      $\{\bm{\beta}_{\mu},\bm{\mathcal{I}}_{\mu}\}_{m}$. The updating order is,
      column-wise, from left to right.}
    \label{tab:gibbs}
      \begin{tabular}{llll}
        \toprule
        Margin component $(1)$ & ...  & Margin component ($M$) & Copula component
                                                                 ($C$)\\
        \midrule
        $(1.1)$ $\{\bm{\beta}_{\mu},\bm{\mathcal{I}}_{\mu}\}_{1}|\{\bm{\beta}_{\mu},\bm
        {\mathcal{I}}_{\mu}\}_{-1}$ & ...  & $(M.1)$
                                             $\{\bm{\beta}_{\mu},\bm{\mathcal{I}}_{\mu}\}_{M}|\{\bm{\beta}_{\mu},\bm
                                             {\mathcal{I}}_{\mu}\}_{-M}$ & $(C.1)$
                                                                           $\{\bm{\beta}_{\lambda},\bm{\mathcal{I}}_{\lambda}\}_{C}|\{\bm{\beta}_{\lambda}
                                                                           ,\bm{\mathcal{I}}_{\lambda}\}_{-C}$\\ $(1.2)$
        $\{\bm{\beta}_{\phi},\bm{\mathcal{I}}_{\phi}\}_{1}|\{\bm{\beta}_{\phi},\bm
        {\mathcal{I}}_{\phi}\}_{-1}$ & ...  & $(M.2)$
                                              $\{\bm{\beta}_{\phi},\bm{\mathcal{I}}_{\phi}\}_{M}|\{\bm{\beta}_{\phi},\bm
                                              {\mathcal{I}}_{\phi}\}_{-M}$ & $(C.2)$
                                                                             $\{\bm{\beta}_{\tau},\bm{\mathcal{I}}_{\tau}\}_{C}|\{\bm{\beta}_{\tau},\bm
                                                                             {\mathcal{I}}_{\tau}\}_{-C}$\\ $(1.3)$
        $\{\bm{\beta}_{\nu},\bm{\mathcal{I}}_{\nu}\}_{1}|\{\bm{\beta}_{\nu},\bm
        {\mathcal{I}}_{\nu}\}_{-1}$ & ...  & $(M.3)$
                                             $\{\bm{\beta}_{\nu},\bm{\mathcal{I}}_{\nu}\}_{M}|\{\bm{\beta}_{\nu},\bm
                                             {\mathcal{I}}_{\nu}\}_{-M}$ & \\ $(1.4)$
        $\{\bm{\beta}_{\kappa},\bm{\mathcal{I}}_{\kappa}\}_{1}|\{\bm{\beta}_{\kappa}
        ,\bm{\mathcal{I}}_{\kappa}\}_{-1}$ & ...  & $(M.4)$
                                                    $\{\bm{\beta}_{\kappa},\bm{\mathcal{I}}_{\kappa}\}_{M}|\{\textbf{}m{\beta}_
                                                    {\kappa}
                                                    ,\bm{\mathcal{I}}_{\kappa}\}_{-M}$ & \\
        \bottomrule
      \end{tabular}
  \end{center}
\end{table}

We also implement a fully Bayesian version of the two-stage estimation method for
comparison purposes. In the two-stage approach, we first independently estimate the
posterior of marginal models and then estimates the posterior of copula model conditional
on the posterior of marginal models. The idea of the two-stage estimation is widely used
in copula modeling, as it reduces the computational complexity in calculating the
posterior for high-dimensional copula models.  We compare the efficiency based on our
proposed joint approach and the two-stage approach in the simulation studies in Section
\ref{sec:simulation} and the real data application in Section \ref{sec:application}.

\subsection{Forecasting evaluation}

We evaluate the model performance based on a $K$-fold out-of-sample log predictive score
(LPS) \citep{geweke2010comparing} and out-of-sample VaR. The LPS is an overall forecasting
evaluation tool based on predictive densities and VaR focuses on tail risks.

We denote the LPS as
\begin{align*}
  \mathrm{LPS}=\frac{1}{K}\sum\nolimits _{k=1}^{K}\log
  p_k({\bm{y}}_{d}|{\bm{y}}_{-d},\bm{x}),
\end{align*}
where ${\bm{y}}_{d}$ is an $n_{d}\times p$ matrix containing $n_{d}$ observations in the
$k$:th testing dataset, and ${\bm{y}}_{-d}$ denotes the training observations used for
estimation. If we assume that the observations are independent conditional on
$\{\bm{\beta},\bm{\mathcal{I}}\}$, then
\begin{align}
  \label{eq:predict-density}
  p_k({\bm{y}}_{d}|{\bm{y}}_{-d},\bm{x})=\int\!\prod\nolimits
  _{i\in{d}}p(\bm{y}_{i}|\{\bm{\beta},\bm{\mathcal{I}}\},\bm{x}_{i})p(\{\bm{\beta
  },\bm{\mathcal{I}}\}|{\bm{y}}_{-d})\mathrm{d}\{\bm{\beta},\bm{\mathcal{I}}\}.
\end{align}
The LPS is easily calculated by averaging
$\prod_{i\in{d}}p(\bm{y}_{i}|\{\bm{\beta},\bm{\mathcal{I}}\},\bm{x}_{i})$ over the
posterior draws from $p(\{\bm{\beta},\bm{\mathcal{I}}\}|{\bm{y}}_{-d})$.  This requires
sampling from each of the $K$ posteriors
$p_k(\{\bm{\beta},\bm{\mathcal{I}}\}|{\bm{y}}_{-d})$ for $k=1,...,K$. The LPS has three main
advantages: i) LPS is based on out-of-sample probability forecasting, which is the
unquestionable model evaluation tool for decision makers \citep{geweke2001bayesian,
  geweke2010comparing}; ii) LPS is easy to compute based on Monte Carlo simulations; and iii)
LPS is not sensitive to the choice of the priors compared with the marginal likelihood
based criterions \citep{kass1993bayes,richardson1997bayesian}.

We also evaluate our model's capacity in capturing tail-dependences based on the
out-of-sample performance of the Value-at-Risk (VaR) estimation.  VaR is a measure
defining how a portfolio $R_{t}$ of assets is likely to decrease over a certain time
period as $p( R_{t} \leq VaR_t ) = \alpha$. This means that we have 100 $( 1 - \alpha )\%$
confidence that the loss in the period $\Delta t$ is not larger than $VaR_t$. In
multivariate financial analysis, VaR is a direct connection to the tail of the
distribution. Therefore, a desired model is expected to produce reasonable and adequate
VaR values in empirical studies.

We follow \citet{huang2009estimating} and consider our portfolio return $R_{ t}$ composed
by a two-asset return denoted as $y_{1 t}$ and $y_{2 t}$. The portfolio return can be
approximately written as:
\begin{align*}
  R_{t}=\omega_1 y_{1t} + \omega_2y_{2t},~\mathrm{with}~ \omega_1+\omega_2=1,
\end{align*}
where $\omega_1$ and $\omega_2$ are the portfolio weights of the two assets. In our work,
we arbitrarily consider the two assets' weights to be equal, but this is not a constraint
generally. There is no analytical solution for calculating $VaR_t$ with our copula
models. We approximate the out-of-sample VaR by firstly simulating the assets vector
$\bm{y}_t=(y_{1t},y_{2t})$ from the joint predictive density
$p(\bm{y}_t|\bm{y}_{1:(t-1)},\bm{x})$ based on Equation (\ref{eq:predict-density}) and
then calculating the portfolio return $R_t$ and its $\% \alpha$ empirical quantile.

\section{Simulation studies}
\label{sec:simulation}

In this section, we evaluate the efficiency of our proposed covariate-dependent copula
model with simulated data. We use exactly the same prior and algorithmic settings for
simulations in this section and the real data application in
Section~\ref{sec:application}.  Our simulation studies apply to the Joe-Clayton copula,
Clayton copula, Gumbel copula, and the Student's \emph{t}-copula. In this section, we
choose to present the results based on the Joe-Clayton copula for
illustration. Comparisons of these four copulas based on real data using different
modeling strategies are presented in Section \ref{sec:application}.

We explain our data generating process (DGP) in Section~\ref{sec:dgp} with combinations of
lower, moderate, and high tail-dependences. We describe the algorithmic details
in Section~\ref{sec:algorithmic-details}. Our model comparisons in Section
\ref{sub:Model-comparison-dgp} are based on out-of-sample evaluation.

\subsection{DGP for copula models with dynamic dependences}
\label{sec:dgp}

We randomly generate bivariate variables $\bm{y}$ with regression-type-of covariate
effects in all copula features ($\lambda_L$, $\lambda_U$) and features of marginal
distributions, such as mean, variance, skewness and kurtosis. The procedure of generating
bivariate random variables with dynamic dependences is different from the usual DGP that
is used in regression.  The detailed DGP settings for parameters in the Joe-Clayton copula
and split-\emph{t} margins are listed in Table \ref{tab:dgp-setting}. We document the
algorithm as follows.

\begin{enumerate}[noitemsep]
\item Generate copula features $\bm{\lambda}_L$ and $\bm{\lambda}_U$.
  \begin{enumerate}
  \item Randomly generate $n$ lower tail-dependence $\bm{\lambda}_L$ and $n$ upper
    tail-dependence $\bm{\lambda}_U$ from beta distribution to allow for dynamic
    dependences, with their mean values $0<\bar {\bm{\lambda}}_L,~\bar {\bm{\lambda}}_U<1$
    and standard deviation values $\sigma_L$ and $\sigma_U$, respectively.
  \item Calculate their linear predictors $\bm{\eta}_L = l(\bm{\lambda}_L)$ and
    $\bm{\eta}_U = l(\bm{\lambda}_U)$ with the link function $l(\cdot)$.
  \end{enumerate}
\item Generate dependent $n\times q$ covariates $\bm{X}$ for copula features
  $\bm{\lambda}_L$ and $\bm{\lambda}_U$.
  \begin{enumerate}
  \item Given a $q\times 1$ coefficient vector $\bm{\beta}$, randomly pick one element
    $\beta_1 \neq 0$ from $\bm{\beta}$ and set the remaining $q-1$ elements as
    $\bm{\beta}_{\{q-1\}}$.
  \item Randomly generate an $n\times (q-1)$ matrix $\bm{X}_{\{q-1\}}$ from $U[0,~1]$.
  \item Calculate $\bm{X}_1=(\bm{\eta}-\bm{X}_{\{q-1\}}\bm{\beta}_{\{q-1\}})/\beta_1$.
  \item Let $\bm{X}=[\bm{X}_{\{q-1\}},\bm{X}_1]$, so that the copula feature $\bm{\lambda}$
    satisfies the covariate-dependent structure $\bm{\lambda}=l^{-1}(\bm{X}\bm{\beta})$.
  \end{enumerate}
\item Generate an $n\times 2$ matrix $\bm{u}$ in $[0, 1]^2$ for the Joe-Clayton copula
  with dependences $\bm{\lambda}_L$ and $\bm{\lambda}_U$ in step 1.(a) by the conditional
  method \citep{nelsen2006introduction}.
\item Generate two sets of split$-t$ marginal features including the mean $\bm{\mu}$, the
  scale $\bm{\phi}$, degrees of freedom $\bm{\nu}$, skewness $\bm{\kappa}$ vectors and the
  covariates $\bm{X}$ followed by step 1-2 with the link function used in
  Table~\ref{tab:prior-setup}.

\item Calculate the $n\times 2$ quantile $\bm{y}$ based on the generated $\bm{u}$ in step
  3 and the marginal model features in step 4.
\end{enumerate}

The covariate matrix $\bm{X}$ generated in Step 2 gives prominence to have a variational
dependence on tail. When $\bm{X}$ is only an $n \times 1$ constant vector, our DGP
generates random variables with static tail-dependences. The coefficient vector
$\bm{\beta}$ indicates the covariate effects on tail-dependences.

\begin{table}
  \begin{center}
    \caption{The DGP settings for features in the Joe-Clayton copula and split-\emph{t}
      margins. The first element of the coefficient vector $\bm{\beta}$ is set as the
      intercept. The DGP settings involves 16 combinations of different mean values of
      $\lambda_L$ and $\lambda_U$ given in the table.}
    \label{tab:dgp-setting}
    \resizebox{\columnwidth}{!}{%
      \begin{tabular}{llllll}
        \toprule
        Feature&Distribution&mean&sd&&True coefficients $\bm{\beta}$ in the
                                       link function $l(\bm{x}\bm{\beta})$\\
        \midrule
        mean $\mu$&Beta&$0$&$1$&&$[1,~1,~-1,~1,~-1,~0,~0]$\\
        scale $\phi$&logNormal&$1$&$1$&&$[1,~1,~-1,~1,~-1,~0,~0]$\\
        degrees of freedom $\nu$&logNormal&$6$&$1$&&$[1,~1,~-1,~1,~-1,~0,~0]$\\
        skewness $\kappa$&logNormal&$1$&$1$&&$[1,~1,~-1,~1,~-1,~0,~0]$\\
        \\
        lower tail-dependence
        $\lambda_L$&Beta&$[0.3,~0.5,~0.7,~0.9]$&$0.1$&&$[1,~1,~-1,~1,~-1,~0,~0,~1,~-1,
                                                        ~1,~-1,~0,~0]$\\
        upper tail-dependence
        $\lambda_U$&Beta&$[0.3,~0.5,~0.7,~0.9]$&$0.1$&&$[1,~1,~-1,~1,~-1,~0,~0,~1,~-1,
                                                        ~1,~-1,~0,~0]$\\
        \midrule
      \end{tabular}
    }
  \end{center}
\end{table}

\subsection{Priors and algorithmic details}
\label{sec:algorithmic-details}

The initial values for coefficients $\bm{\beta}$ in the MCMC are obtained by numerically
optimizing the posterior distribution from a two-stage approach with a randomly selected
10\% of the training dataset. This is to guarantee a finite log posterior
value in the first MCMC iteration. The initial variable selection indicators
$\bm{\mathcal{I}}$ are set to one in our application.

The prior for variable selection indicators is $Bern(0.5)$, which means a variable has an
equal probability to be included or excluded in the model. The priors for coefficients are
listed in Table \ref{tab:prior-setup} and their details are described in Section
\ref{sec:prior}. Changing the mean and variance of ``Prior belief on feature'' in Table
\ref{tab:prior-setup} would affect the implied prior settings on the intercepts $\beta_0$,
as well as the posterior. From our experiments, the changes in LPS are always smaller than
one when we double or halve the mean and variance values of ``Prior belief on feature''.

The scaling hyperparameters in Section \ref{sec:prior-slop-vs} are one in the priors of
all $\bm{\beta}_s$. Within each step of the Metropolis-Hastings algorithm, the proposal
distribution for $\bm{\beta}$ is a multivariate Student's \emph{t} distribution with its
mean based on three-step Newton iterations and six degrees of freedom.  The proposal for
variable selection indicators is $Bern(0.5)$, as described in Section \ref{sec:mcmc}.  Our
proposed MCMC algorithm allows for variable selection to reduce model complexity. In
high-dimensional problems, one may set a stricter proposal to exclude more unnecessary
variables.

\begin{table}
  \begin{center}
    \caption{The prior belief on features and the implied priors for $\beta_0$,
      $\bm{\beta}_s$ and $\mathcal{\bm{I}}$ used in the covariate-dependent copula
      models.}
    \label{tab:prior-setup}
    \resizebox{\columnwidth}{!}{%
      \begin{tabular}{p{0.25\textwidth}lll}
        \toprule
        Feature & Link function & Prior belief on feature & $p(\beta_{0})$\\
        \midrule
        mean $\mu$&{identity}&$N(\mathrm{mean}=0,~\mathrm{var}=1)$&$N(\mathrm{mean}=0,
                                                                    ~\mathrm{var}=1)$\\
        scale
        $\phi$&{log}&$\mathrm{logNorm}(\mathrm{mean}=1,~\mathrm{var}=1)$&$N(\mathrm
                                                                          {mean}=-0.34,\mathrm{var}=0.69)$\\
        degrees of freedom
        $\nu$&{log}&$\mathrm{logNorm}(\mathrm{mean}=5,~\mathrm{var}=10)$&$N(\mathrm
                                                                          {mean}=1.44,~\mathrm{var}=0.34)$\\
        skewness $\kappa$&{log}&$\mathrm{logNorm}(\mathrm{mean}=1,~\mathrm{var}=1)$&
                                                                                     $N(\mathrm{mean}=-0.34,\mathrm{var}=0.69)$\\
        \\
        Kendall's $\tau$ \& tail-dependences $\lambda_L$,
        $\lambda_U$&{logit}&$\mathrm{Beta}(\mathrm{mean}=0.2,~\mathrm{var}=0.05)$&$N
                                                                                   (\mathrm{mean}=-2.55,~\mathrm{var}=6.91)$\\
        \midrule
        \midrule
        $p(\mathcal{I}_j) \sim Bern(0.5)$&  & &\\
        $p(\bm{\beta}_{s}|\mathcal{\bm{I}}) \sim \bm{N}(\bm{0},~\bm{\Sigma})$ &where
                                                                                $\bm{\Sigma} = diag(1,..,1)$ &&\\

        \bottomrule
      \end{tabular}
    }
  \end{center}

\end{table}

The efficiency of the MCMC is monitored via the inefficiency factor
$\text{IF}=1+2\sum_{i=1}^{\infty}\rho_{i}$, where $\rho_{i}$ is the autocorrelation at lag
$i$ in the MCMC iterations. The inefficiency factor for all parameters in our studies are
below $40$.

An R package 
for estimating the covariate-dependent copula model with our MCMC scheme, posterior
summary and visualization, DGP, and model comparison is
available 
upon request. The program runs on a Linux cluster node with two $2.0$ GHz Intel Xeon CPUs
with total of 16 cores and $128$ GB RAM. Parallelization is used when cross-validation is
applied. We set the number of MCMC iterations to be $20,000$, and discard the first $10\%$
as burn-in. It takes roughly two hours to run a four-fold cross-validation with the DGP
dataset based on the full model, including a covariate-dependent structure and variable
selection in all parameters.

\subsection{Forecasting comparisons}
\label{sub:Model-comparison-dgp}

In this paper, the notation \emph{CD.} means covariate-dependent structure is applied to
copula features, \emph{VS.} means variable selection in copula features, and \emph{Const.}
means only constants are used when modeling copula features.

We firstly check whether a different copula parametrization for the Joe-Clayton copula
influences the forecasting performance. To achieve this, two types of datasets are
generated. Datasets of the first type are generated with the covariate-dependent structure
only applied to copula features (as shown in Table \ref{tab:dgp-setting}) to eliminate the
effect from marginal models, while datasets of the second type are generated without
covariate-dependent structure in any model features.  We estimate the reparameterized
Joe-Clayton copula with three parameterizations $(\lambda_L,\lambda_U)$,
$(\lambda_L,\tau)$ and $(\lambda_U,\tau)$ for these two types of datasets. Table
\ref{tab:bb7-parameterization-test} shows the LPS results for moderate lower
tail-dependence and moderate upper tail-dependence with their mean values $0.3$ and $0.5$
and standard deviation $0.1$. With our proposed joint approach, the LPS values for all
three reparameterized models are consistent when the true model is covariate-independent,
which is reasonable because all the relationships among the three features are
deterministic. When the lower tail-dependence and upper tail-dependence in the true DGP
model are covariate-dependent, we do not know which covariates would affect the other two
parameterizations $(\lambda_L,\tau)$ and $(\lambda_U,\tau)$ because the covariates are
generated with the parameterization $(\lambda_L,\lambda_U)$. Therefore, we use all the
covariates available to model the two reparameterized models. The variation in LPS values
is larger than the covariate-independent case, which is as expected. For exactly the same
settings but with variable selection applied to the covariates, the deviation is greatly
mitigated, as shown in Table~\ref{tab:bb7-parameterization-test}. For the same
parameterization, we notice that the LPS does not differ significantly using either the
independent Bernoulli prior or the uniform prior for variable selection indicators. The
results in the remainder of the paper are based on the Bernoulli prior.

\begin{table}
  \centering
  \caption{LPS comparison based on four-fold cross-validation on different
    parameterizations for Joe-Clayton copula and prior settings for variable selection
    indicators.  Each dataset consists of $1,000$ observations with given mean values
    $\bar \lambda_L=0.3$ and $\bar \lambda_U=0.5$, and standard deviation $0.1$. To
    eliminate the effects from marginal models, the marginal models are from
    split-\emph{t} densities with the same feature settings as in Table
    \ref{tab:dgp-setting} with all slopes $\bm{\beta}_s$ being zero. We only estimate the
    intercepts in marginal features. }
  \label{tab:bb7-parameterization-test}
  \begin{tabular}{llrrr}
    \toprule
    &&\multicolumn{3}{c}{Parameterization}\\
    \cline{3-5}
    True DGP dependence& Estimation
                         strategy&$(\lambda_L,\lambda_U)$&$(\lambda_L,\tau)$&$(\lambda_U,\tau)$\\
    \midrule
    \emph{Const.}&\emph{Const.}& $-984.62$&$-985.71$&$-985.56$\\
    \cline{2-5}
    \emph{CD.} &\emph{CD.+VS.+Bern Prior}&$-984.76$&$-986.45$&$-989.34$\\
    &\emph{CD.+VS.+Uniform Prior}&$-983.39$&$-985.02$&$-988.22$\\
    &\emph{CD.}&$-983.54$&$-1303.63$&$-1117.77$\\
    \bottomrule
  \end{tabular}
\end{table}

We now look into how the covariate-dependent structure improves forecasting performances
for both the joint modeling strategy and the widely used two-stage approach.  We generate
16 datasets of size
$n=1,000$ with a combination of different mean values in the lower tail-dependence ($\bar
\lambda_L$) and mean values in the upper tail-dependence ($\bar
\lambda_U$), which are shown in Table~\ref{tab:dgp-setting}.  Under each DGP setting, we
estimate the model with our proposed joint approach and a two-stage approach.  For each
approach, we estimate our model under two different settings: i) our full
covariate-dependent structure plus a variable selection scheme; and ii) simply modeling
the constants in the copula features. As such, four modeling strategies are applied for
each dataset. It is worth mentioning that we use a full Bayesian method for estimating
both the marginal models and copula component in the two-stage approach. The copula
component inference in the second-stage is based on the full posterior in marginal models,
which is a significantly improved version of the typical two-stage approach where only the
parameter mean of the marginal model is used.

Table~\ref{tab:simulation} shows the LPS of a four-fold cross-validation for the
Joe-Clayton copula. Note that the LPS is not comparable among different datasets with
different lower tail-dependence and upper tail-dependence settings. For all of the 16 DGP
settings, the Joe-Clayton copula with full covariates in the copula structure in a joint
modeling approach performs the best compared to the other three strategies. The
significance level, with regard to the LPS differences, increases as the strength of the
tail-dependence increases. Under the same model setting, the joint approach is always
better than the two-stage approach. In the joint approach, with covariates applied to
copula features, there is an improvement in the out-of-sample performance, which is
significant especially when a moderate or high tail-dependence occurs.

\begin{table}
  \begin{center}
    \caption{LPS comparison based on four-fold cross-validation for Joe-Clayton copula
      with 16 DGP settings where the true model is covariate-dependent as described in
      Table \ref{tab:bb7-parameterization-test}.  A total of 64 simulations are carried
      out based on different combination of lower tail-dependence and upper
      tail-dependence. Each dataset consists of $1,000$ observations with given mean
      values $\bar \lambda_L$ and $\bar \lambda_U$, and standard deviation $0.1$. We
      estimate the Joe-Clayton copula model with split-\emph{t} margins based on four
      modeling strategies (\emph{Joint + CD.+VS.}, \emph{Joint + Const.}, \emph{Two-stage+
        CD.+VS.} and \emph{Two-stage + Const.}).  }
    \label{tab:simulation}
    \resizebox{\columnwidth}{!}{%
      \begin{tabular}{rlccccccccccc}
        \toprule
        DGP
        settings&&\multicolumn{2}{c}{$\bar\lambda_U=0.3$}&&\multicolumn{2}{c}{
                                                            $\bar\lambda_U=0.5$}
        &&\multicolumn{2}{c}{$\bar\lambda_U=0.7$}&&\multicolumn{2}{c}{$\bar\lambda_U=0
                                                    .9$}\\
        \cline{3-4}\cline{6-7}\cline{9-10}\cline{12-13}
                &MCMC&\emph{CD.+VS.}&\emph{Const.}&&\emph{CD.+VS.}&\emph{Const.}&&\emph{CD.+VS.
                                                                                   }&\emph{Const.}&&\emph{CD.+VS.}&\emph{Const.}\\
        \midrule
        $\bar\lambda_L=0.3$&\emph{Joint}&$\mathbf{-519.56}$&$-520.91$&&$\mathbf
                                                                        {-506.90}$&$-508.95$&&$\mathbf{-427.72}$&$-432.35$&&$\mathbf{-273.93}$&$-306.99
                                                                                                                                                $\\
                &\emph{Two-stage}&$-523.25$
        &$-522.00$&&$-510.60$&$-511.75$&&$-444.32$&$-439.68$&&$-310.67$&$-321.38$\\
        \\
        $\bar\lambda_L=0.5$&\emph{Joint}&$\mathbf{-501.33}$&$-502.57$&&$\mathbf{-468.30
                                                                        }$&$-471.97$&&$\mathbf{-424.30}$&$-436.54$&&$\mathbf{-244.02}$&$-268.56$\\
                &\emph{Two-stage}&$-510.51$&$-507.29$&&$-476.68$&$-474.30$&&$-446.38$&$-451.83$
                                                  &&$-299.08$&$-314.36$\\
        \\
        $\bar\lambda_L=0.7$&\emph{Joint}&$\mathbf{-440.81}$&$-454.16$&&$\mathbf{-424.20
                                                                        }$&$-439.24$&&$\mathbf{-380.30}$&$-390.38$&&$\mathbf{-243.16}$&$-244.78$\\
                &\emph{Two-stage}&  $-457.76$
        &$-460.83$&&$-440.01$&$-440.70$&&$-397.72$&$-402.37$&&$-283.96$&$-295.11$\\

        \\
        $\bar\lambda_L=0.9$&\emph{Joint}&$\mathbf{-228.83}$&$-256.11$&&$\mathbf{-218.61
                                                                        }$&$-294.52$&&$\mathbf{-241.21}$&$-255.13$&&$\mathbf{-210.11}$&$-269.86$\\
                &\emph{Two-stage}&$-244.01$&-294.00&&$-292.74$&$-317.60$&&$-280.67$&$-289.88$&&
                                                                                                $-259.15$&$-297.25$\\
        \bottomrule
      \end{tabular}
    }
  \end{center}

\end{table}

We further check the efficiency of variable selection for mitigating a misspecified model
when unnecessary variables are included in the lower and upper tail-dependences.  We
repeat the DGP with the settings described in Table~\ref{tab:dgp-setting} but set all the
slope coefficients to be zero, i.e. the copula features in the true model are not affected
by the covariates. Then we estimate a Joe-Clayton copula model with three settings: i)
artificially generated covariate structure applied to the lower and upper tail-dependences
with joint modeling approach plus variable selection; ii) same as i) but without variable
selection; and iii) only the intercepts are included in copula features, which is the
benchmark model. The LPS values in Table \ref{tab:simulation-vs-comparison} shows that
including unnecessary covariates in the tail-dependence features could decrease a model's
forecast capacity. But this can be alleviated via the Bayesian variable selection
scheme. Our simulation studies indicate the average efficiency improvement due to variable
selection is 32.5\% for different combinations of lower and upper tail-dependences.

\begin{table}
  \begin{center}
    \caption{LPS comparison based on four-fold cross-validation for Joe-Clayton copula
      with 16 DGP settings where the true model is covariate-independent in all
      Joe-Clayton copula and split-\emph{t} marginal features.  A total of 48 simulations
      are conducted based on different combination of lower tail-dependence and upper
      tail-dependence. Each dataset consists of $1,000$ observations with given mean
      values $\bar \lambda_L$ and $\bar \lambda_U$, and standard deviation $0.1$. We
      estimate the Joe-Clayton copula model jointly with three modeling strategies
      (\emph{CD.+VS.}, \emph{CD.} and \emph{Const.}).}
    \label{tab:simulation-vs-comparison}
    \resizebox{\columnwidth}{!}{%
      \begin{tabular}{rcccccccccccccccc}
        \toprule
        DGP
        settings&&\multicolumn{3}{c}{$\bar\lambda_U^{(DGP)}=0.3$}&&\multicolumn{3}{c}{
                                                                    $\bar\lambda_U=0.5$}
        &&\multicolumn{3}{c}{$\bar\lambda_U=0.7$}&&\multicolumn{3}{c}{$\bar\lambda_U=0
                                                    .9$}\\
        \cline{3-5}\cline{7-9}\cline{11-13}\cline{15-17}\\
                &&\emph{CD.+VS.}&\emph{CD.}&\emph{Const.}&&\emph{CD.+VS.}&\emph{CD.}&\emph
                                                                                      {Const.}
                                           &&\emph{CD.+VS.}&\emph{CD.}&\emph{Const.}&&\emph{CD.+VS.}&\emph{CD.}&\emph
                                                                                                                 {Const.}\\
        \midrule
        $\bar\lambda_L=0.3$&&$-980.97$&$-988.00$&$-960.95$&&$-984.03$&$-989.06$&$-967
                                                                                 .98$&&$-973.88$&$-978.02$&$-965.53$&&$-991.62$&$-996.06$&$-957.56$\\
        $\bar\lambda_L=0.5$&&$-959.95$&$-964.87$&$-958.32$&&$-994.70$&$-998.60$&$-975
                                                                                 .69$&&$-968.10$&$-977.24$&$-967.35$&&$-992.07$&$-998.23$&$-962.59$\\
        $\bar\lambda_L=0.7$&&$-975.20$&$-975.58$&$-968.03$&&$-974.75$&$-981.30$&$-968
                                                                                 .03$&&$-970.17$&$-973.98$&$-958.12$&&$-987.26$&$-991.37$&$-970.30$\\
        $\bar\lambda_L=0.9$&&$-985.44$&$-989.74$&$-964.16$&&$-980.20$&$-982.21$&$-967
                                                                                 .38$&&$-985.02$&$-992.05$&$-965.61$&&$-971.10$&$-975.07$&$-969.30$\\
        \midrule

        \multicolumn{15}{l}{$\mathrm{Average~efficiency~improvement~due~to~variable
        ~selection:~} \frac{1}{16}\sum
        \nolimits_{i=1}^{16}(\mathrm{LPS}_{\mathrm{CD.+VS.}}(i)-\mathrm{LPS}_{\mathrm
        {CD.}}(i))/(\mathrm{LPS}_{\mathrm{Const.}}(i)-\mathrm{LPS}_{\mathrm{CD.}}(i))
        =32.5\%$} \\
        \bottomrule
      \end{tabular}
    }
  \end{center}

\end{table}

We measure the forecasting accuracy of tail-dependence using the out-of-sample mean
absolute error (MAE) loss function.  We calculate the out-of-sample MAE loss for the
results in Table \ref{tab:simulation} based on a four-fold cross-validation.  For all four
modeling strategies (\emph{Joint+CD.+VS.}, \emph{Joint+Const.}, \emph{Two-stage+CD.+VS.}
and \emph{Two-stage+Const.}), the average values of the out-of-sample MAE for the lower
and upper tail-dependences with respect to the 16 simulations are presented in Table
\ref{tab:loss}.  Our proposed \emph{Joint+CD.+VS.} approach has the lowest average MAE in
both lower and upper tail-dependences compared to the other three modeling
strategies. This approach significantly improves the accuracy of lower tail dependence
forecasting, which is also the focus in empirical analysis.

\begin{table}
  \caption{Mean values based on four-fold cross-validation for the out-of-sample mean
    absolute error (MAE) loss for a total of 16 simulations in Table~\ref{tab:simulation}
    with four modeling strategies (\emph{Joint + CD.+VS.}, \emph{Joint + Const.},
    \emph{Two-stage+ CD.+VS.} and \emph{Two-stage + Const.}). }
  \label{tab:loss}
  \begin{center}
    \begin{tabular}{cccccc}
      \toprule
      &
        \emph{Joint+CD.+VS.}&\emph{Joint+Const.}&\emph{Two-stage+CD.+VS.}&\emph{Two-stage+Const.} \\
      \midrule
      $\lambda_L$ & $0.084$ & $0.091$ & $0.114$ &$0.094$ \\
      $\lambda_U$ & $0.082$ & $0.083$ & $0.105$ & $0.092$ \\
      \bottomrule
    \end{tabular}
  \end{center}
\end{table}

\begin{figure}
  \centering \includegraphics[width=0.7\textwidth]{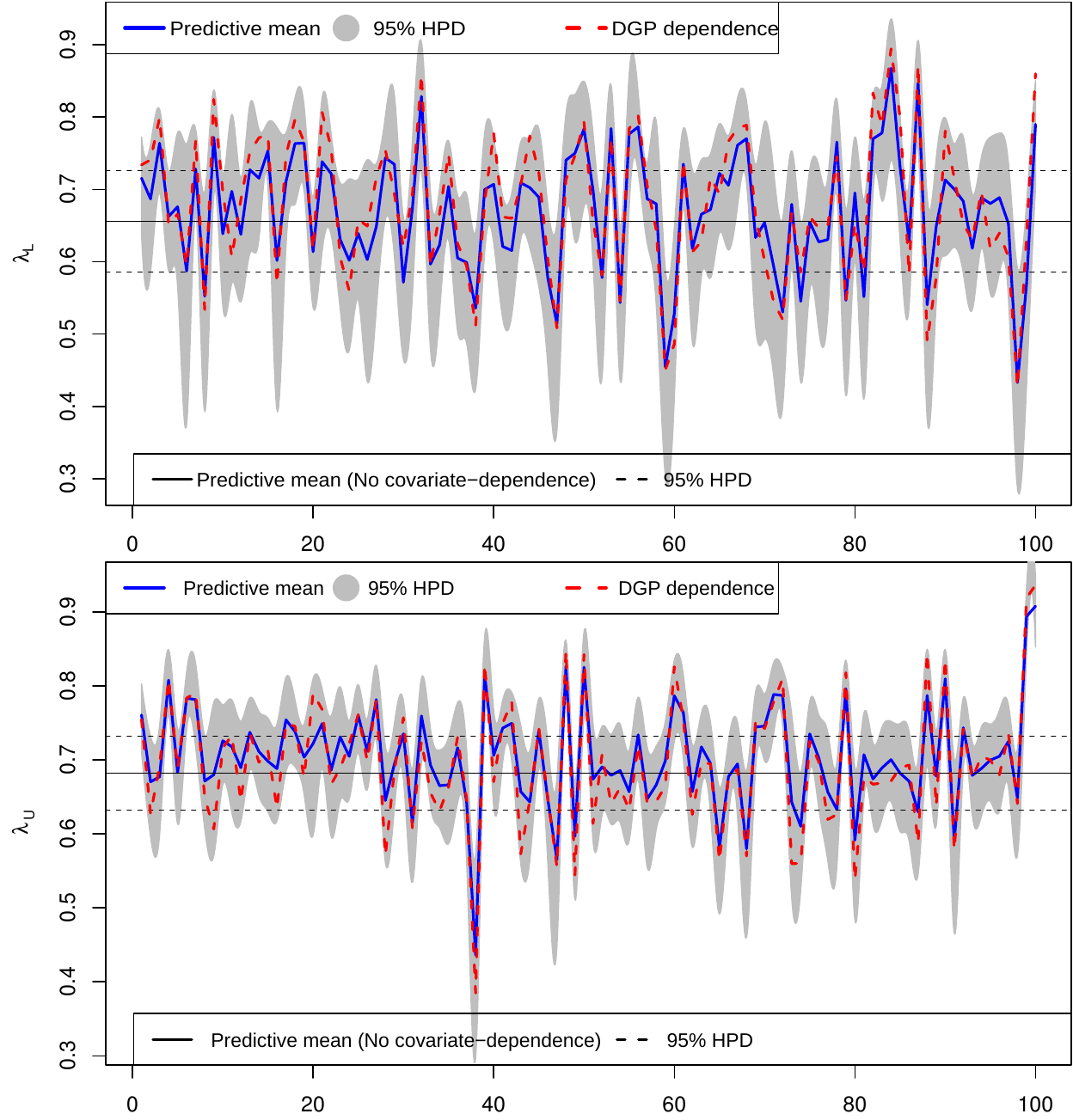}
  \caption{The excerpt summary of the predictive lower tail-dependences (top) and upper
    tail-dependences (bottom). The predictive lower and upper tail-dependences are
    calculated from the four-fold cross-validation of two modeling strategies
    (\emph{Joint+CD.+VS.} and \emph{Joint+Const.}). The data consists of $1,000$
    observations with $\bar {\bm{\lambda}}_L=0.7$ and $\bar {\bm{\lambda}}_U=0.7$ as noted
    in Table~ \ref{tab:simulation}. The plot only depicts predictive lower and upper
    tail-dependences for an excerpt of the first 100 observations for better
    visualization.}
  \label{fig:post-summary-dgp}
\end{figure}

In the end of this section, we present the summary of predictive copula features in
Figure~\ref{fig:post-summary-dgp} for the joint modeling approach with i) a
covariate-dependent structure applied to copula features (\emph{Joint+CD.+VS.})  and ii)
only constant estimated in copula features (\emph{Joint+Const.}).
Figure~\ref{fig:post-summary-dgp} is based on a DGP with
$\bar {\bm{\lambda}}_L=0.7$ and $\bar {\bm{\lambda}}_U=0.7$.  We choose to present this
setup because in a real financial application, as shown in Section~\ref{sec:application},
this setting is close to the estimated tail dependences during financial crisis
periods. The two subplots in Figure~\ref{fig:post-summary-dgp} depict the modeling
performances by comparing their capacities to recover the true DGP dependences.  This
shows that our joint modeling approach with covariate-dependent structure applied to
copula features nicely captures the true DGP dependences even when very high dependence
occurs in the last part of the second subplot. The model with only constants estimated in
the copula features only captures the mean of true DGP dependences, but fails to cover the
volatility of the dependences even with a 95\% highest probability density (HPD).

\section{Application to financial data forecasting}
\label{sec:application}

In order to illustrate our method, we apply it to a financial application with daily stock
returns. The copula models are reparameterized Joe-Clayton copula, Clayton copula, Gumbel
copula, and Student's \emph{t}-copula with split-\emph{t} distributions on the continuous
margins. For the discrete case, see the latent variables approach for the Gaussian copula
\citep{pitt2006efficient} and the extension to a general copula
\citep{smith2012estimation}.

\subsection{The S\&P 100 and S\&P 600 data}
\label{sec:SP100-SP600}

\begin{figure}
  \centering \includegraphics[width=\textwidth]{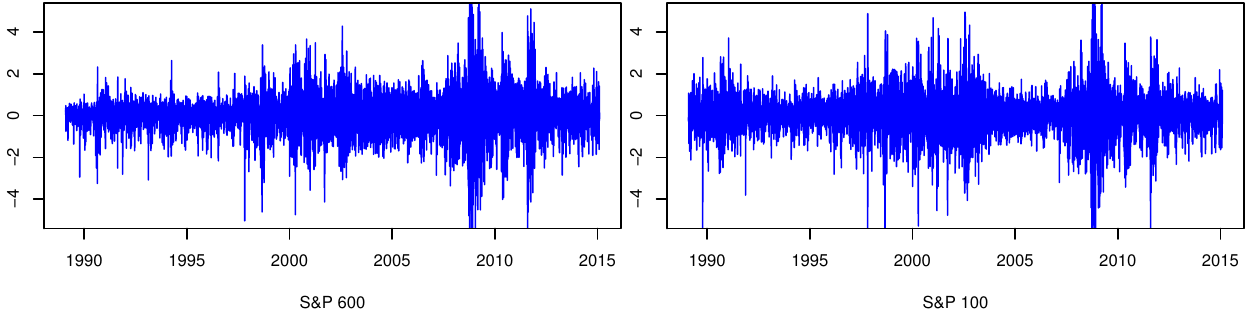}
  \caption{The daily returns of the S\&P 600 and S\&P 100 indices from February 01, 1989
    to February 06, 2015.}
  \label{fg:SP100-SP600}
\end{figure}

Our data are daily returns from the S\&P 100 and S\&P 600 daily stock market indices
during the period from February 01, 1989 to February 06, 2015 (see Figure
\ref{fg:SP100-SP600}). The S\&P 100 index includes the largest and most established
companies in the U.S. and is a subset of the well-known S\&P 500 index. The S\&P 600 index
covers the small capitalization companies which present the possibility of greater capital
appreciation, but at greater risk. The S\&P 600 index covers roughly three percent of the
total US equities market.

\begin{figure}
  \centering \includegraphics[width=0.7\textwidth]{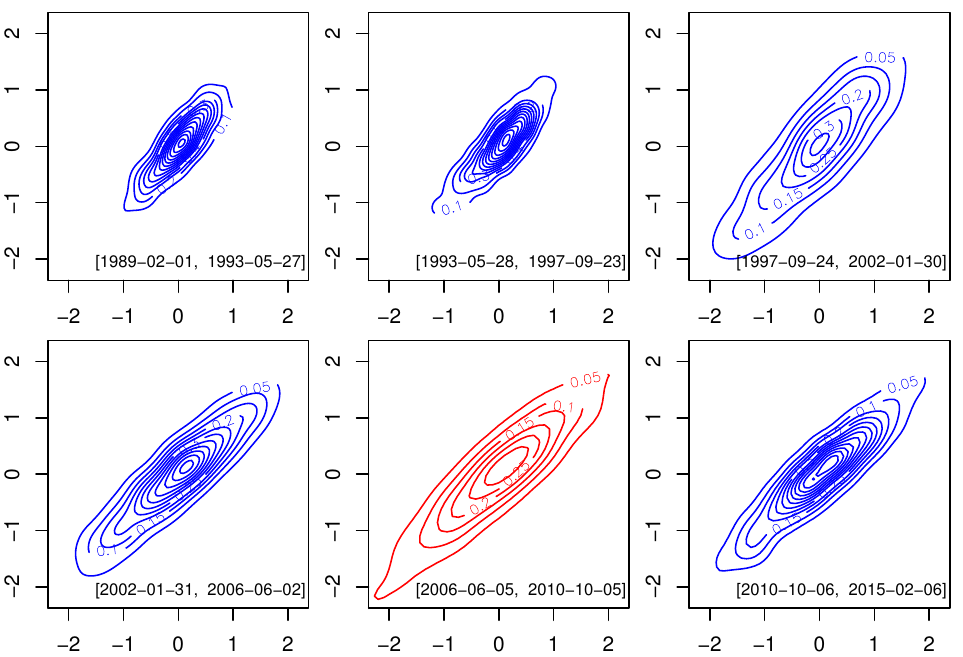}

  \caption{Contour plots for bivariate empirical densities with daily returns of the S\&P
    100 (y-axis) and the S\&P 600 (x-axis) indices from February 01, 1989 to February 06,
    2015. The full data are partitioned into six data blocks in chronological order
    (marked in squared brackets) and the densities are estimated via a normal kernel
    method within the corresponding blocks. The subplot in the middle of the second row is
    for the period during the 2008 financial crisis. }
  \label{fg:SP100-SP600-empCopulaDensity}
\end{figure}

Figure \ref{fg:SP100-SP600} shows the time series of the daily returns.  It is shown that
there is huge volatility in the returns for both the S\&P 100 and the S\&P 600 during the
2008 financial crisis. Figure~\ref{fg:SP100-SP600-empCopulaDensity} depicts six empirical
densities for the \textsf{Return} variable that are partitioned into blocks estimated with
normal kernel methods by assuming independent observations. The empirical densities in
Figure~\ref{fg:SP100-SP600-empCopulaDensity} show that the dependences, especially the
lower tail-dependence, changes over time. There are signs of extreme positive dependences
during the 2008 financial crisis. Figure~\ref{fg:SP100-SP600-empCopulaDensity} also
suggests that there is a strong lower tail-dependence and upper tail dependence during and
after the 2008 financial crisis. The two-parameter Joe-Clayton copula is appropriate for
modeling the data without rotation. Furthermore, \citet{joe2005asymptotic} shows
that the usual two-stage approach for copula estimation is not fully efficient for extreme
dependences, which is the case in the empirical studies exhibited in
Figure~\ref{fg:SP100-SP600-empCopulaDensity}.

The covariates described in Table \ref{tab:sp-covariates} in the margins and in the copula
features are from nonlinear transformation of historical returns and prices of the
series. We standardize each covariate to have zero mean and unit variance. The covariates
can be classified into two groups, time-varying variables (\textsf{RM1}, \textsf{RM5} and
\textsf{RM20}) and volatility indicators (\textsf{CloseAbs95}, \textsf{CloseAbs80},
\textsf{MaxMin95}, \textsf{MaxMin80}, \textsf{CloseSqr95} and \textsf{CloseSqr80}). The
reason we use time-varying variables as covariates for modeling tail-dependences is that
the empirical study of \citet{patton2012copula} shows that there exists significant
time-varying dependence between the S\&P 100 and the S\&P 600. Volatility indicators are
also used as covariates for measuring tail-dependences because heteroscedasticity is found
to be a major source of tail-dependences in stock indices (see the discussion and empirical
study in \citet{poon2003extreme} and \citet{hilal2011hedging}). \citet{poon2003extreme}
also argues that the choice of volatility indicators is unimportant. Therefore, one can
replace the volatility indicators with any other volatility variables, and thus our
modeling scheme can be generally applied.  The volatility indicators are originally used
in \citet{geweke2007smoothly}.  \citet{villani2009regression} and \citet{li2010flexible}
apply similar covariates to univariate regression density estimation on the S\&P 500 data
using mixtures of Gaussian and split-\emph{t} densities, and they show that those
covariates can efficiently capture the volatility in stock indices.

\begin{table}
  \begin{center}
    \caption{Description of covariates used in the model for the S\&P 100 and S\&P 600
      data.}
    \label{tab:sp-covariates}
    \begin{tabular}{lp{0.8\textwidth}l}
      \toprule
      Covariate & Description \\
      \midrule
      \textsf{Return}  & Daily return $y_{t}=100\log(p_{t}/p_{t-1})$ where $p_{t}$ is
                         the closing price. \\
      \textsf{RM1} & Return of last day. \\
      \textsf{RM5} & Return of last week. \\
      \textsf{RM20} & Return of last month. \\
      \textsf{CloseAbs95} & Geometrically decaying average of absolute returns
                            $(1-\rho)\sum\nolimits _{s=0}^{\infty}\rho^{s}|y_{t-2-s}|$
                            with $\rho=0.95$. \\
      \textsf{CloseAbs80} & Geometrically decaying average of past absolute returns
                            with $\rho=0.80$. \\
      \textsf{MaxMin95} & Measure of volatility $(1-\rho)\sum\nolimits
                          _{s=0}^{\infty}\rho^{s}(\log(p_{t-1-s}^{h})-\log(p_{t-1-s}^{l}))$
                          with $\rho=0.95$, where $p^{h}$ and $p^{l}$ are the highest and lowest prices.
      \\
      \textsf{MaxMin80} & Measure of volatility with $\rho=0.80$. \\
      \textsf{CloseSqr95} & Geometrically decaying average of returns
                            $((1-\rho)\sum\nolimits _{s=0}^{\infty}\rho^{s}y_{t-2-s}^{2})^{1/2}$ with
                            $\rho=0.95$. \\
      \textsf{CloseSqr80} & Geometrically decaying average of returns with
                            $\rho=0.80$. \\
      \bottomrule
    \end{tabular}
  \end{center}
\end{table}

\subsection{Posterior summary for S\&P 100 and S\&P 600 data}

\begin{figure}
  \centering \hspace{0.3cm} \includegraphics[width=0.48\textwidth]{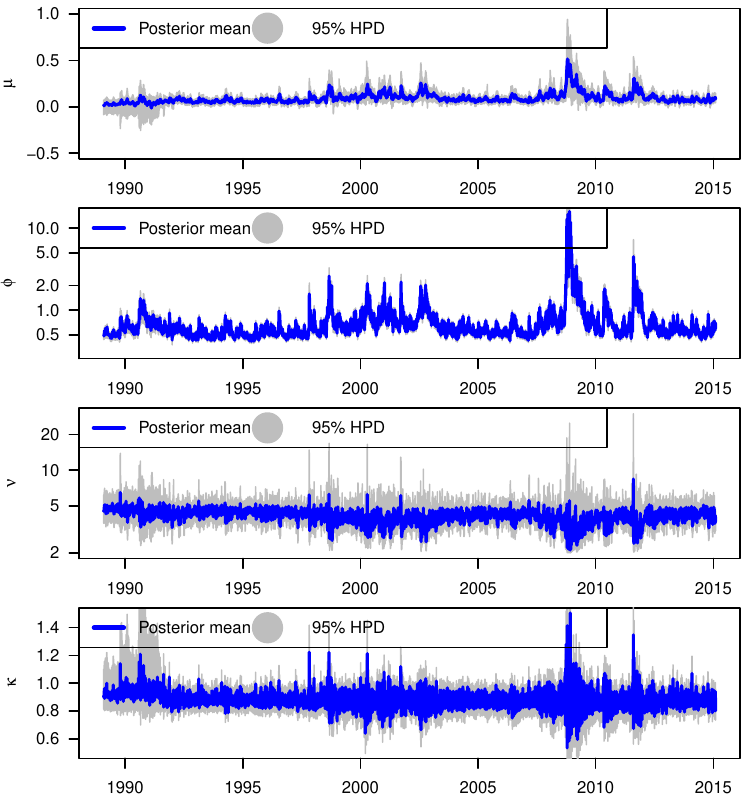}
  \includegraphics[width=0.48\textwidth]{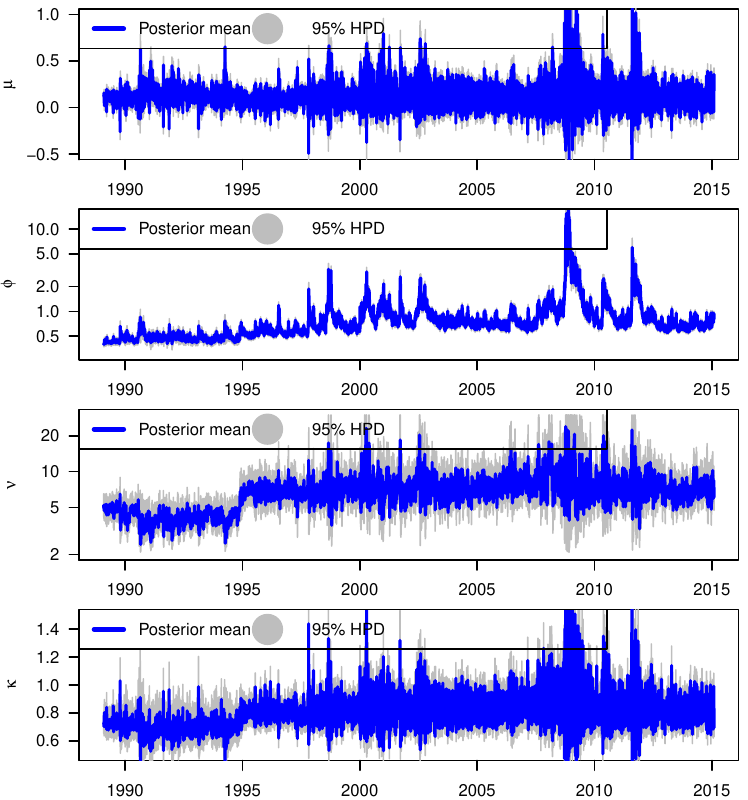}\\
  \includegraphics[width=\textwidth]{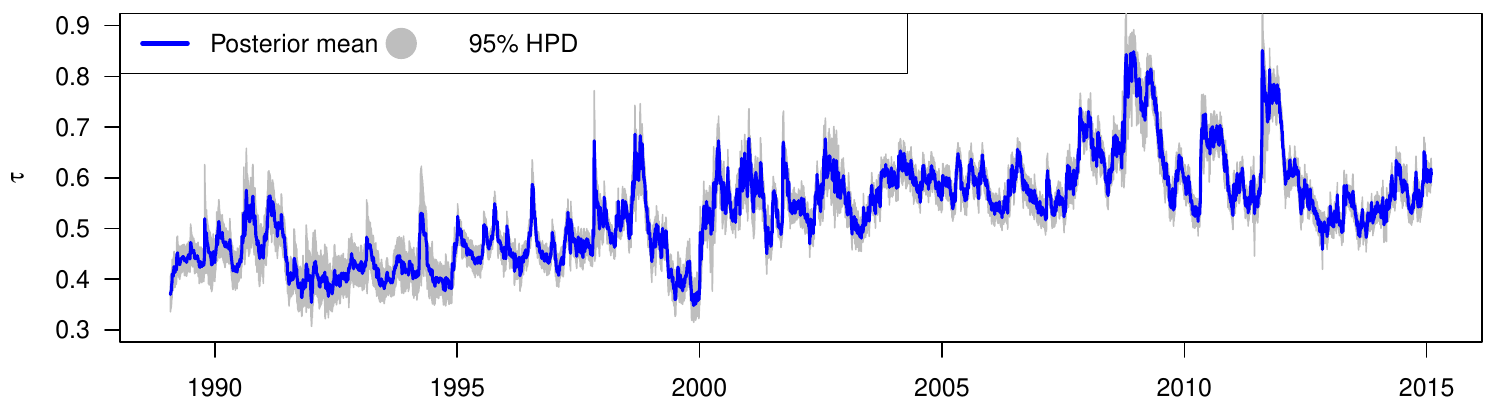}\\
  \includegraphics[width=\textwidth]{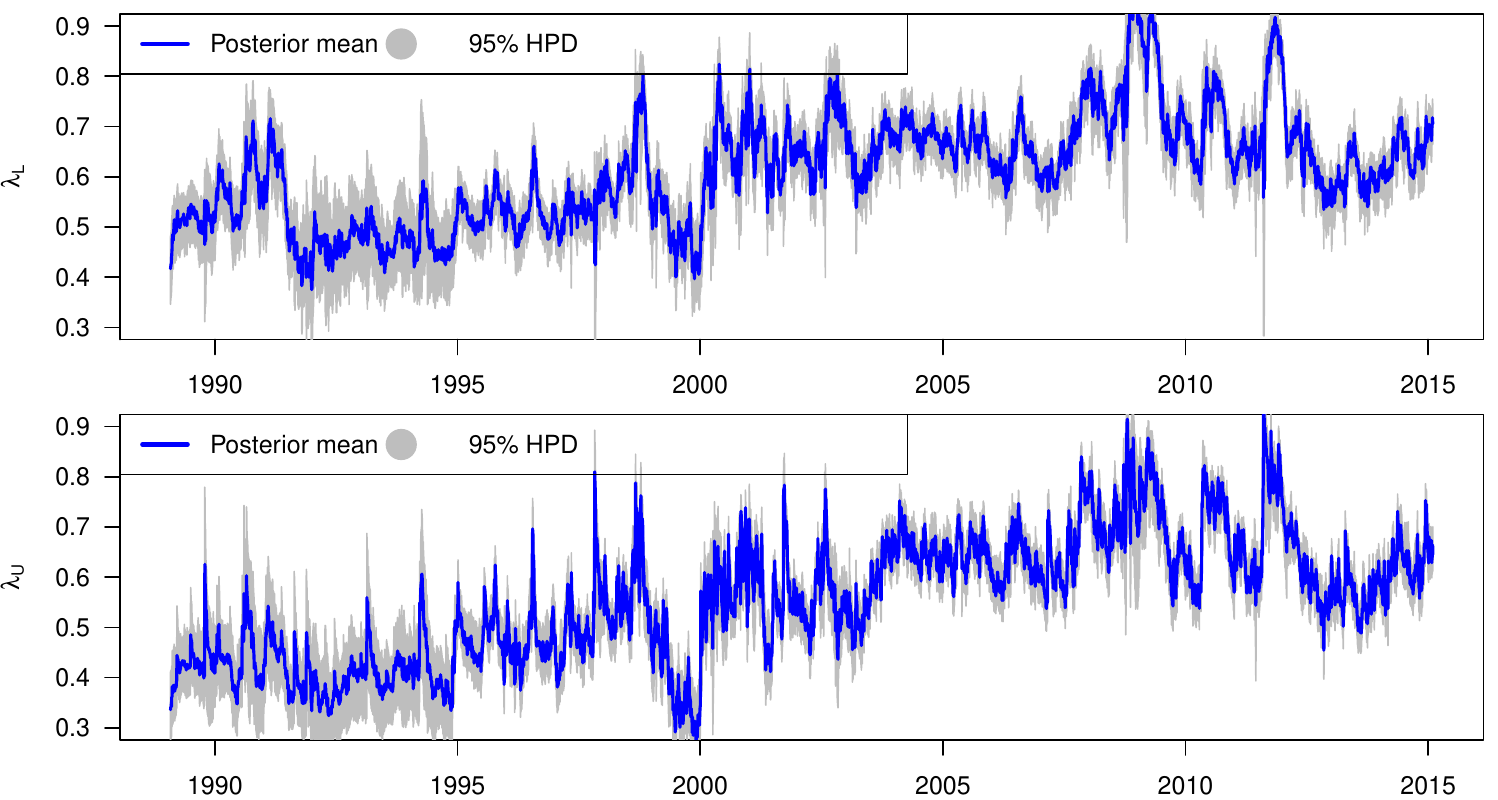}
  \caption{The posterior mean and the 95\% highest probability density (HPD) intervals for
    the Joe-Clayton copula model reparameterized in terms of lower tail-dependence
    $\lambda_L$ and Kendall's $\tau$ for the S\&P 600 and the S\&P 100 data from February
    01, 1989 to February 06, 2015. The top-left and top-right four subplots show time
    series of the mean, scale (in log scale), degrees of freedom (in log scale) and
    skewness features in the margins of the S\&P 600 and the S\&P 100, respectively. The
    three subplots at the bottom are the time series for Kendall's $\tau$, the lower
    tail-dependence, and the upper tail-dependence $\lambda_U$, which is derived based on
    the posterior of lower tail-dependence and Kendall's $\tau$.}
  \label{fg:SP100-SP600-post}

\end{figure}

We firstly present the posterior summary for the Joe-Clayton copula model for S\&P 100 and
S\&P 600 data with split-\emph{t} margins in two parameterizations in terms of
$(\lambda_L,~\lambda_U)$ in Table~\ref{tab:sp-fullrun}, and $(\lambda_L,~\tau)$ in
Figure~\ref{fg:SP100-SP600-post}.  The mean posterior acceptance probability is $71\%$.
Our MCMC algorithm reduces to the standard Metropolis-Hastings algorithm by setting the
number of Newton's iterations as zero.The MCMC scheme provides a higher posterior
acceptance probability and a lower inefficiency factor compared to standard
Metropolis-Hastings, where the average acceptance probability is below $20\%$ in our
model.

Our marginal models are similar to the model used by \citet{li2010flexible}, except that
the mean of the split-\emph{t} is fixed in their model. We focus on explaining the results
in the copula component, and one can refer to the work of \citet{li2010flexible} for a
detailed interpretation of the marginal models.  The variable selection results show that
important variables for the lower tail-dependence are \textsf{RM5} from the S\&P 100,
\textsf{CloseAbs95} from the S\&P 600, \textsf{CloseAbs80} from both margins,
\textsf{MaxMin95} from the S\&P 100, and \textsf{CloseSqr95} from both margins.  Variables
with large posterior inclusion probabilities in the upper-tail dependence are
\textsf{CloseAbs95} from the S\&P 100, \textsf{CloseSqr95} from the S\&P 100 and
\textsf{CloseSqr80} from the S\&P 600.

The same covariates in the S\&P 100 and S\&P 600 tend to be highly correlated. The
posterior is therefore hard to evaluate using a poorly constructed MCMC algorithm because
the likelihood is weakly identified when the same type of covariates in each margin are
included in the the copula features.  Table \ref{tab:sp-fullrun} shows that when a
covariate in one margin is selected in the copula feature, the same covariate in the other
margin does not appear in the copula with only two exceptions. This indicates that our
variable selection algorithm is efficient in removing superfluous covariates. Also note
that the initial values for variable selection indicators are all one. That is, initially,
all covariates are included in the model. Our inclusion probabilities for variable
selection indicators in Table~\ref{tab:sp-fullrun} are very selective in the sense that
very few indicators are close to an undecided region around $0.5$, and all of the finally
selected variables do not rely on the initially included variables.

\begin{table}
  \begin{center}
    \caption{The summary for the posterior mean values of the Joe-Clayton copula model
      reparameterized in terms of lower tail-dependence ($\lambda_L$) and upper
      tail-dependence ($\lambda_U$) for the S\&P 100 and the S\&P 600 data from February
      01, 1989 to February 06, 2015. In the results for the copula component, the first
      and second rows for $\beta$ and $\mathcal{I}$ are obtained from the combined
      covariates that are used in the first and second marginal models. The intercepts are
      always included in the model. Variables that are selected with an inclusion
      probability greater than $0.5$ are marked in bold. The average of inefficiency
      factors for all parameters is $29$. The average acceptance probabilities are all
      above $0.70$ within each iteration of the Metropolis-Hastings algorithm.}
    \label{tab:sp-fullrun}
    \resizebox{\columnwidth}{!}{%
      \begin{tabular}{rrrrrrrrrrr}
        \toprule
        &\textsf{Intercept}&\textsf{RM1}&\textsf{RM5}&\textsf{RM20}
        &\textsf{CloseAbs95}& \textsf{CloseAbs80}& \textsf{MaxMin95}
        & \textsf{MaxMin80}&\textsf{CloseSqr95}& \textsf{CloseSqr80}\\

        \cmidrule(r){2-11}
        \multicolumn{11}{c}{Marginal component (S\&P 600)}\\

        $\beta_{\mu}$&$0.146$&$\mathbf{0.120}$&$0.000$&$0.006$&$0.001$&$\mathbf{0.116}$
                                                 &$-0.001$&$0.002$&$-0.013$&$\mathbf{-0.046}$\\
        $\mathcal{I}_{\mu}$&$1.00$&$\mathbf{1.00}$&$0.01$&$0.22$&$0.02$&$\mathbf{0.96}$
                                                 &$0.05$&$0.08$&$0.24$&$\mathbf{0.53}$\\

        $\beta_{\phi}$&$-0.318$&$\mathbf{0.041}$&$\mathbf{-0.115}$&$-0.004$&$-0.010$&$
                                                                                      -0.012$&$\mathbf{0.217}$&$0.002$&$0.019$&$\mathbf{0.243}$\\
        $\mathcal{I}_{\phi}$&$1.00$&$\mathbf{1.00}$&$\mathbf{1.00}$&$0.16$&$0.13$&$0.17
                                                                                   $&$\mathbf{0.94}$&$0.12$&$0.18$&$\mathbf{0.96}$\\

        $\beta_{\nu}$&$1.405$&$0.008$&$\mathbf{-0.230}$&$0.041$&$-0.160$&$-0.098$&$-0
                                                                                   .220$&$\mathbf{-0.414}$&$\mathbf{0.428}$&$\mathbf{0.641}$\\
        $\mathcal{I}_{\nu}$&$1.00$&$0.13$&$\mathbf{0.78}$&$0.27$&$0.48$&$0.38$&$0.50$&
                                                                                       $\mathbf{0.65}$&$\mathbf{0.56}$&$\mathbf{0.74}$\\

        $\beta_{\kappa}$&$-0.248$&$\mathbf{-0.116}$&$0.000$&$0.000$&$-0.039$&$-0.009$&$
                                                                                       -0.057$&$-0.015$&$\mathbf{0.174}$&$-0.003$\\
        $\mathcal{I}_{\kappa}$&$1.00$&$\mathbf{1.00}$&$0.02$&$0.17$&$0.29$&$0.16$&$0.38
                                                                                   $&$0.32$&$\mathbf{0.83}$&$0.13$\\

        \cmidrule(r){2-11}
        \multicolumn{11}{c}{Marginal component (S\&P 100)}\\

        $\beta_{\mu}$&$0.091$&$0.000$&$-0.001$&$\mathbf{-0.019}$&$0.010$&$0.000$&$0.011
                                                                                  $&$-0.005$&$0.011$&$0.020$\\
        $\mathcal{I}_{\mu}$&$1.00$&$0.02$&$0.05$&$\mathbf{0.57}$&$0.30$&$0.13$&$0.25$&
                                                                                       $0.15$&$0.19$&$0.26$\\

        $\beta_{\phi}$&$-0.432$&$-0.001$&$\mathbf{-0.109}$&$\mathbf{-0.026}$&$\mathbf{0
                                                                              .481}$&$\mathbf{-0.191}$&$\mathbf{0.135}$&$0.045$&$\mathbf{-0.477}$&$\mathbf{0
                                                                                                                                                   .339}$\\
        $\mathcal{I}_{\phi}$&$1.00$&$0.05$&$\mathbf{1.00}$&$\mathbf{0.60}$&$\mathbf{0
                                                                            .96}$&$\mathbf{0.96}$&$\mathbf{0.58}$&$0.33$&$\mathbf{1.00}$&$\mathbf{0.98}$\\

        $\beta_{\nu}$&$0.641$&$-0.024$&$\mathbf{-0.173}$&$0.067$&$-0.151$&$-0.160$&$-0
                                                                                    .001$&$0.035$&$-0.005$&$-0.055$\\
        $\mathcal{I}_{\nu}$&$1.00$&$0.20$&$\mathbf{0.62}$&$0.36$&$0.41$&$0.42$&$0.34$&
                                                                                       $0.43$&$0.32$&$0.27$\\

        $\beta_{\kappa}$&$-0.140$&$\mathbf{-0.058}$&$0.000$&$-0.005$&$0.033$&$0.000$&$0
                                                                                      .020$&$0.000$&$-0.077$&$0.007$\\
        $\mathcal{I}_{\kappa}$&$1.00$&$\mathbf{1.00}$&$0.05$&$0.15$&$0.20$&$0.09$&$0.21
                                                                                   $&$0.02$&$0.38$&$0.08$\\

        \cmidrule(r){2-11}
        \multicolumn{11}{c}{Copula component ($C$)}\\

        $\beta_{\lambda_L}$&$0.508$&$0.003$&$-0.017$&$0.000$&$\mathbf{1.497}$&$\mathbf{
                                                                               -0.449}$&$0.031$&$-0.030$&$\mathbf{0.359}$&$0.040$\\
        &&$0.001$&$\mathbf{-0.072}$&$0.002$&$0.088$&$\mathbf{0.268}$&$\mathbf{-0.811}$&
                                                                                        $0.043$&$\mathbf{-0.498}$&$-0.090$\\

        $\mathcal{I}_{\lambda_L}$&$1.00$&$0.07$&$0.42$&$0.00$&$\mathbf{1.00}$&$\mathbf
                                                                               {0.84}$&$0.27$&$0.20$&$\mathbf{0.75}$&$0.37$\\
        &&$0.04$&$\mathbf{0.55}$&$0.04$&$0.44$&$\mathbf{0.64}$&$\mathbf{0.96}$&$0.16$&
                                                                                       $\mathbf{0.82}$&$0.45$\\

        $\beta_{\lambda_U}$&$0.246$&$-0.001$&$-0.001$&$0.000$&$0.044$&$0.001$&$1.384$&
                                                                                       $0.005$&$0.199$&$\mathbf{0.347}$\\
        &&$-0.001$&$-0.005$&$-0.005$&$\mathbf{-0.887}$&$-0.066$&$-0.466$&$0.226$&
                                                                                  $\mathbf{-0.229}$&$-0.157$\\

        $\mathcal{I}_{\lambda_U}$&$1.00$&$0.03$&$0.02$&$0.11$&$0.11$&$0.08$&$1.00$&$0
                                                                                    .13$&$0.47$&$\mathbf{0.94}$\\
        &&$0.02$&$0.09$&$0.36$&$\mathbf{1.00}$&$0.41$&$0.43$&$0.43$&$\mathbf{0.67}$&$0
                                                                                     .38$\\
        \bottomrule
      \end{tabular}
    }
  \end{center}
\end{table}

Figure \ref{fg:SP100-SP600-post} represents the posterior of marginal features from
split-\emph{t} and copula features, Kendall's $\tau$ correlation, lower tail-dependence
and the derived upper tail dependences between the S\&P 100 and the S\&P 600, and their
95\% HPD. Both the lower and upper tail-dependences are not strong during low volatility
periods, but they vary over time significantly. A very high dependence could occur in the
tail even though the overall Kendall's $\tau$ correlation is relatively small.

\begin{figure}
  \centering \includegraphics[width=\textwidth]{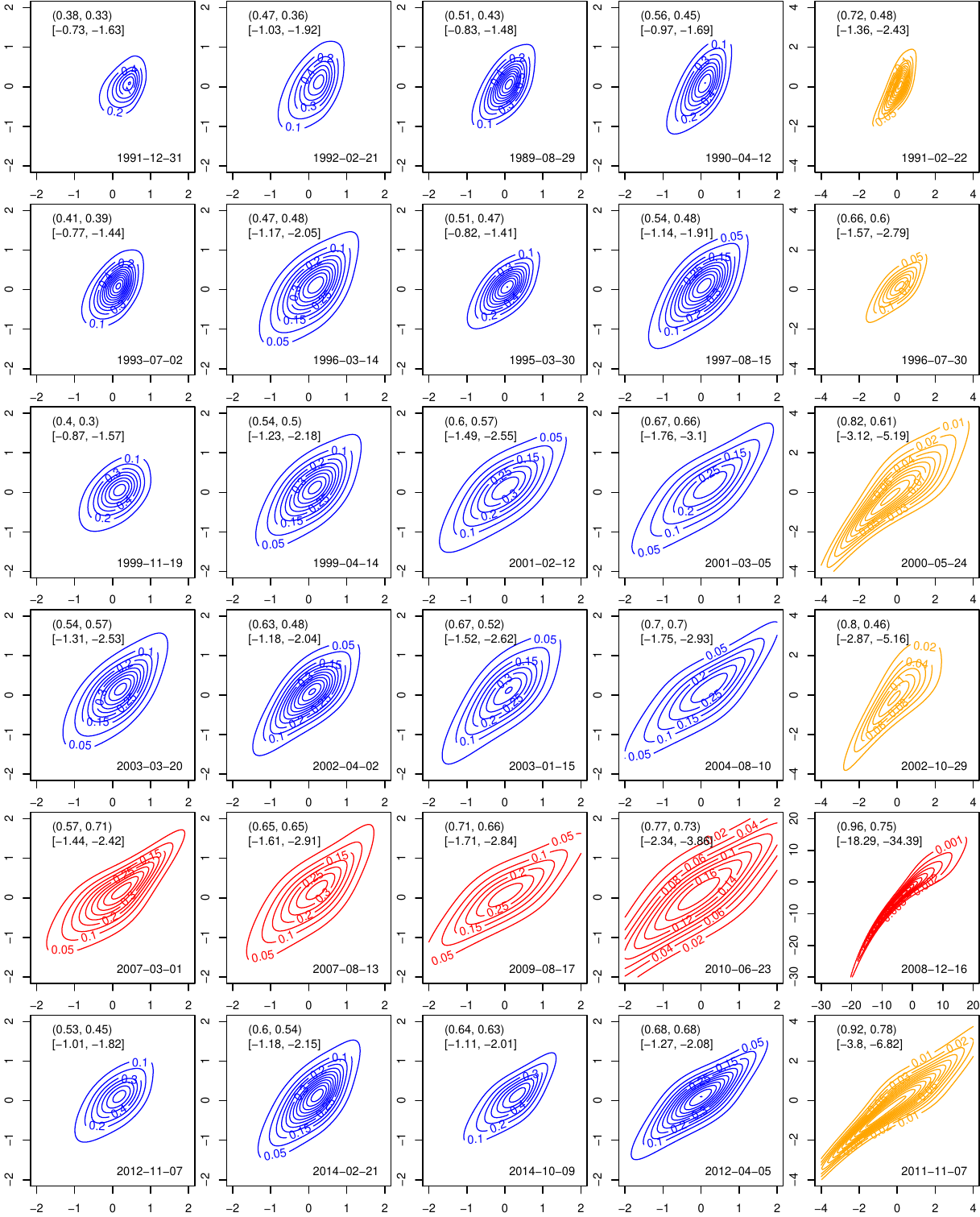}
  \caption{Contour plots of the posterior densities for selected dates with the
    Joe-Clayton copula model reparameterized in terms of lower tail-dependence $\lambda_L$
    and upper tail-dependence $\lambda_U$ for the S\&P 100 (y-axis) and the S\&P 600
    (x-axis) data from February 01, 1989 to February 06, 2015. The posterior means for
    $\lambda_L$ and $\lambda_U$ are marked in round brackets. The 5\% VaR and 1\% VaR are
    presented in square brackets. Five days (column-wise view) sorted by ascending order
    of lower tail-dependences (1\%, 25\%, 50\%, 75\% and 99\% quantiles) are selected in
    each data block from the six blocks (row-wise view) used in Figure
    \ref{fg:SP100-SP600-empCopulaDensity}.  The subplots in the second last row
    (highlighted in red) are for the period during the 2008 financial crisis. The
    magnitudes are enlarged in the subplots of the last column due to the extremely high
    dependences.}
  \label{fig:contour-post}
\end{figure}

Figure \ref{fig:contour-post} depicts the contour plot for the posterior density of the
Joe-Clayton copula model. The dates selected are based on the ascending order of the
estimated lower tail-dependences. We can see that the lower tail-dependence is much higher
during the crisis in comparison to the dates before. Also note from the shape of the
contour lines that the empirical densities in Figure~\ref{fg:SP100-SP600-empCopulaDensity}
overestimate the dependences during the non-crisis period and underestimate the
tail-dependences during the 2008 financial crisis because they assume temporally
independent observations. Our covariate-dependent copula model is capable of capturing the
dynamic nature of tail dependences.

\subsection{Forecasting comparisons}
\label{sub:Model-comparison-sp}

In our time series application, we calculate the LPS based on the posterior estimation of
80\% of the historical data and the density forecasting of the last 20\% of the data. It
evaluates the quality of the one-step-ahead predictions using
\begin{align*}
  \mathrm{LPS}&=\log p(\bm{y}_{(T+1):(T+p)}|\bm{y}_{1:T})\\
              &=\sum\limits _{i=1}^{p}\log\int
                p(\bm{y}_{T+i}|\{\bm{\beta},\bm{\mathcal{I}}\},\bm{y}_{1:(T+i-1)})p(\{\bm{\beta
                },\bm{\mathcal{I}}\}|\bm{y}_{1:(T+i-1)})\mathrm{d}\{\bm{\beta},\bm{\mathcal{I}
                }\},
\end{align*}
where $\bm{y}_{a:b}$ is the dataset from time $a$ to $b$ and
$\{\bm{\beta},\bm{\mathcal{I}}\}$ are the model parameters. This way of calculating LPS is
usually computationally costly because every one-step-ahead prediction needs a new
posterior sample from the posterior based on the data available at the time of the
forecast. We approximate the LPS by assuming that the posterior does not change much as we
add a few data points to the estimation sample. Each time we predict 10\% of the testing
data based on previously estimated posterior by assuming that the observations are
independent and then update the posterior. Thus calculating LPS for the whole testing data
involves ten times of such a window expanding procedure.  \citet{villani2009regression}
documents that this type of approximation is accurate in the application of a smooth
mixture of Gaussians for density predictions of the S\&P 500 data. Furthermore, with
simple algebra, it can be shown that the global LPS in a copula model equals to the sum of
the LPS values in each marginal model and the copula component (i.e.,
$\mathrm{LPS}=\sum\nolimits _{i=1}^M \mathrm{LPS}_i + \mathrm{LPS}_C$). This allows us to
parallel compute the LPS values and compare the contributions of different components.

Table \ref{tab:lpds} shows the out-of-sample comparisons based on LPS of the Joe-Clayton
copula, Clayton copula, Gumbel copula and Student's \emph{t}-copula in combination with
split-\emph{t}, GARCH(1,1) with normal innovations and SV margins.  The specification of
the univariate SV model and its MCMC details are described in
\citet{kastner2014ancillarity}.  We apply three modeling strategies (\emph{CD.+VS.},
\emph{CD.}, and \emph{Const.}) for each reparameterized copula.  Both the joint modeling
approach and the two-stage approach are applied when possible.  For each combination, the
global LPS for the full model is decomposed into marginal LPS components ($M_1$ and $M_2$)
and a copula LPS component ($C$).

In the joint approach comparison, the Joe-Clayton copula with covariate-dependent structure
and variable selection scheme (\emph{CD.+VS.}) outperforms the other three copulas that
also have the same modeling strategy. The Student's \emph{t}-copula with a correlation
feature and degrees of freedom modeled with covariate-dependent structure is the second
best model, but still falls behind by 36 LPS points.  Compared to only modeling constant
copula features (\emph{Const.}), introducing covariate-dependent copula modeling
(\emph{CD.})  improves forecasting performances regarding the global LPS in the four
copulas. The improvement is further enhanced when the variable selection scheme is added
to the covariate-dependent copula structure (\emph{CD.+VS.}). Among the four copulas, the
Student's \emph{t}-copula is the weakest one for modeling tail-dependences in theory
because its tail dependences rely on small degrees of freedom. Allowing
covariate-dependent structure applied to the degrees of freedom greatly increases the LPS
by about $90$ points from $703.96$ to $792.14$.

In the two-stage approach comparison, we compare three types of margins, including the
default split-\emph{t}, GARCH(1,1) and SV margins with a combination of the four copula
models with covariate-dependent structure applied to all the copula features.  We also
calculate the LPS for bivariate DCC-GARCH and bivariate volatility models as benchmarks
shown at the bottom of Table \ref{tab:lpds}. The Joe-Clayton copula with split-\emph{t}
margins still performs the best. For the same copula, variable selection significantly
improves the forecasting performance in terms of global LPS mainly due to its capacity of
preventing overfitting. The GARCH margin and the SV margin fall behind with distinctions
in terms of the LPS.  Although the bivariate DCC-GARCH does a much better job than
bivariate volatility model, with a gap of $269$ LPS points and the bivariate DCC-GARCH
also outperforms all of the four copulas with an SV margin and GARCH(1,1) margin, our four
covariate-dependent copula models with split-\emph{t} margins using variable selection
outperform the DCC-GARCH with stronger significance of more than $350$ LPS points.

To summarize, copula models with covariate-dependent structure could greatly improve the
capacity of capturing tail dependences, and thus they outperform the commonly used
volatility models. Joint modeling is more efficient than the two-stage approach. Variable
selection is useful in preventing overfitting, which also helps improve the model's
forecasting performances.

\begin{table}
  \begin{center}
    \caption{Model comparison based on LPS. Note that in principle, the global LPS should
      be equal to the summation of LPS values in the marginal and copula components.  But
      for numerical reasons in MCMC, they are not exactly the same. The numerical standard
      errors are all below one in all LPS values.  }
    \label{tab:lpds}
    \resizebox{\columnwidth}{!}{%
      \begin{tabular}{llrrrrrrrrrrrrrrr}
        \toprule

        &&\multicolumn{14}{c}{Reparameterized Copulas}\\
        \cline{3-17}
        &&\multicolumn{3}{c}{Joe-Clayton}&&\multicolumn{3}{c}{Clayton}
                                         &&\multicolumn{3}{c}{Gumbel}&&\multicolumn{3}{c}{Student's \emph{t}-Copula}\\
        \cline{3-5} \cline{7-9} \cline{11-13} \cline{15-17}
        &&\multicolumn{1}{c}{\emph{CD.+VS.}}&\multicolumn{1}{c}{\emph{CD.}}
                                         &\multicolumn{1}{c}{\emph{Const.}}&&\multicolumn{1}{c}{\emph{CD.+VS.}}
                                                                     &\multicolumn{1}{c}{\emph{CD.}}&\multicolumn{1}{c}{\emph{Const.}}
                                            &&\multicolumn{1}{c}{\emph{CD.+VS.}}&\multicolumn{1}{c}{\emph{CD.}}
                                                                           &\multicolumn{1}{c}{\emph{Const.}}&&\multicolumn{1}{c}{\emph{CD.+VS.}}
                                                                                                    &\multicolumn{1}{c}{\emph{CD.}}&\multicolumn{1}{c}{\emph{Const.}}\\
        Margins&\multicolumn{1}{c}{LPS}&&&&&&&&&&&&&&&\\

        \midrule

        &&\multicolumn{15}{c}{(\emph{Joint modeling approach})}\\

        SPLIT-\emph{t}&$M_1$
        &$-1743.12$&$-1745.96$&$-1747.99$&&$-1741.04$&$-1805.79$&$-1789.27$&&$-1754.36$
                                                                           &$-1756.09$&$-1756.21$&&$-1741.47$&$-1859.05$&$-1782.37$\\
        &$M_2$
        &$-1435.98$&$-1430.96$&$-1434.22$&&$-1468.25$&$-1589.48$&$-1561.14$&&$-1485.68$
                                                                           &$-1549.39$&$-1563.01$&&$-1430.07$&$-1516.65$&$-1658.09$\\
        &$C$     &$837.50  $&$832.63  $&$779.14  $&&$690.22  $&$774.34  $&$646.57
                                                                           $&&$797.78  $&$794.91  $&$684.60  $&&$792.14  $&$594.60$  &$703.96  $\\
        &$Global$&$-2344.12$&$-2377.01$&$-2411.06$&&$-2523.75$&$-2639.05$&$-2714.69$&&$
                                                                                       -2448.14$&$-2518.48$&$-2639.58$&&$-2380.12$&$-2786.45$&$-2736.49$\\
        \midrule
        &&\multicolumn{15}{c}{(\emph{Two-stage modeling approach})}\\
        SPLIT-\emph{t}&$M_1$
        &$-1740.10$&$-1737.20$&$-1734.57$&&$-1741.05$&$-1735.22$&$-1735.81$&&$-1737.73$
                                                                           &$-1738.03$&$-1736.72$&&$-1741.47$&$-1735.03$&$-1736.18$\\
        &$M_2$
        &$-1428.39$&$-1428.68$&$-1426.78$&&$-1436.63$&$-1425.18$&$-1426.63$&&$-1427.83$
                                                                           &$-1425.19$&$-1428.41$&&$-1433.41$&$-1431.07$&$-1427.53$\\
        &$C$     &$819.63  $&$225.30  $&$121.84  $&&$694.84  $&$262.49  $&$121.50
                                                                           $&&$781.39  $&$621.78  $&$129.60  $&&$788.22  $&$553.50  $&$289.14   $\\
        &$Global$&$-2346.61$&$-2941.80$&$-3045.04$&&$-2483.93$&$-2903.64$&$-3043.71$&&$
                                                                                       -2392.13$&$-2545.14$&$-3036.39$&&$-2389.41$&$-2617.19$&$-2883.14$\\
        \\
        GARCH(1,1)&$M_1$
        &$-1948.07$&$-1948.07$&$-1948.07$&&$-1948.07$&$-1948.07$&$-1948.07$&&$-1948.07$
                                                                           &$-1948.07$&$-1948.07$&&$-1948.07$&$-1948.07$&$-1948.07$\\
        &$M_2$
        &$-1673.85$&$-1673.85$&$-1673.85$&&$-1673.85$&$-1673.85$&$-1673.85$&&$-1673.85$
                                                                           &$-1673.85$&$-1673.85$&&$-1673.85$&$-1673.85$&$-1673.85$\\
        &$C$     &$702.35  $&$495.35  $&$294.18  $&&$530.48  $&$450.42  $&$148.83
                                                                           $&&$810.39  $&$441.49  $&$147.90  $&&$791.55  $&$632.48  $&$598.06  $\\
        &$global$&$-2919.57$&$-3126.56$&$-3327.73$&&$-3091.44$&$-3171.50$&$-3473.09$&&$
                                                                                       -2811.53$&$-3180.43$&$-3474.01$&&$-2830.37$&$-2989.44$&$-3023.86$\\
        \\
        SV        &$M_1$
        &$-2166.90$&$-2159.29$&$-2178.93$&&$-2154.18$&$-2170.90$&$-2194.82$&&$-2168.17$
                                                                           &$-2162.75$&$-2168.62$&&$-2179.36$&$-2186.61$&$-2183.65$\\
        &$M_2$
        &$-1811.36$&$-1809.61$&$-1814.96$&&$-1844.57$&$-1808.54$&$-1787.42$&&$-1808.61$
                                                                           &$-1828.60$&$-1824.77$&&$-1808.24$&$-1830.06$&$-1826.25$\\
        &$C$     &$964.37  $&$768.19  $&$344.22  $&&$698.30  $&$513.081 $&$126.46
                                                                           $&&$1012.10 $&$733.96  $&$231.85 $&&$1053.19 $&$906.58  $&$755.63  $\\
        &$Global$&$-3013.90$&$-3200.70$&$-3649.67$&&$-3300.46$&$-3466.36$&$-3855.78$&&$
                                                                                       -2964.68$&$-3257.39$&$-3761.53$&&$-2934.40$&$-3110.09$&$-3254.27$\\

        \midrule
        &&\multicolumn{15}{c}{(\emph{Bivariate volatility models})}\\

        \multicolumn{2}{l}{Bivariate DCC-GARCH}&$-2730.78$\\
        \multicolumn{2}{l}{Bivariate SV}&$-2999.63$&\\
        \bottomrule
      \end{tabular}
    }
  \end{center}

\end{table}

\subsection{Value-at-Risk comparisons}
\label{sub:VaR}

\begin{figure}[h!]
  \centering \includegraphics[width=\textwidth]{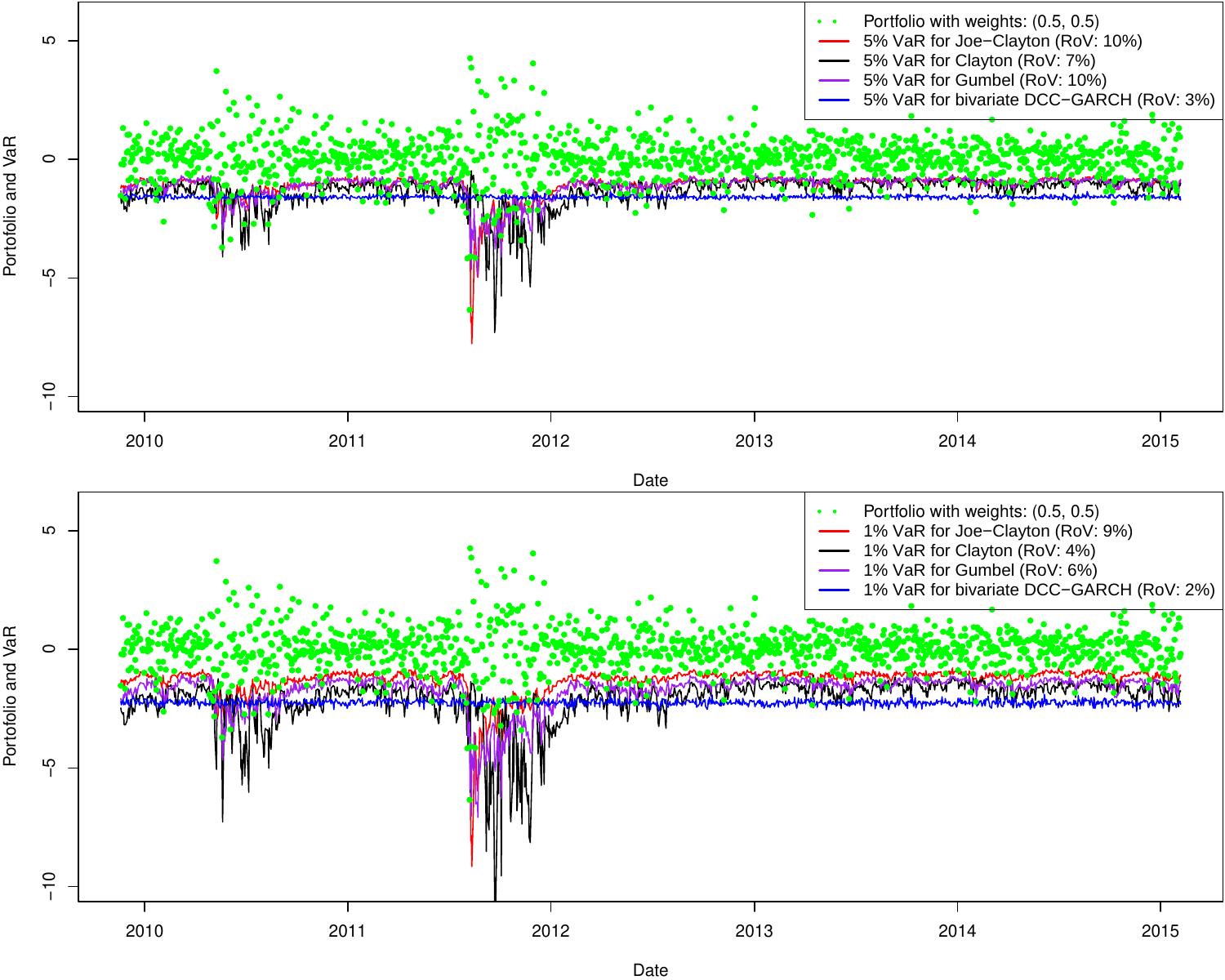}
  \caption{Out-of-sample Value-at-Risk (VaR) for the portfolio of S\&P 100 and S\&P 600
    with equal weights.  The 5\% (top) and 1\% (bottom) VaR are calculated for Joe-Clayton
    copula, Clayton copula, Gumbel copula, and bivariate DCC-GARCH models based on 80\% of
    the historical data. The ratio of violations (RoV) indicating the percentage of sample
    observations lying out of the critical values is also reported for each model.}
  \label{fig:VaR-plot}
\end{figure}

We also compare the out-of-sample Value-at-Risk for the portfolio of S\&P100 and S\&P 600
returns with equal weights based on the reparameterized covariate-dependent Joe-Clayton
copula, Clayton copula, Gumbel copula models, and the bivariate DCC-GARCH model based on
80\% of the historical data in Figure \ref{fig:VaR-plot}. The ratio of violations (RoV) in
Figure \ref{fig:VaR-plot} is the percentage of sample observations lying out of the VaR
critical values. Overall, the Joe-Clayton copula captures the extrema most successfully in
a time-varying fashion and detects more violations compared to the other methods. The
benchmark DCC-GARCH model has the smallest variation in both $5\%$ VaR and $1\%$ VaR
compared to other three models, and also has the smallest RoV, which is not suitable to
reflect the risk of portfolio returns with time-varying effects.  Compared with the
Clayton copula and the Gumbel copula, the difference of VaRs in most regions is not
obvious except for the volatility periods of mid 2010 and late 2011. In these regions, the
$5\%$ VaR and $1\%$ VaR for Gumbel copula overreact to the portfolio's volatility.  From
the $5\%$ VaR plot in the top panel of Figure~\ref{fig:VaR-plot}, we notice that the
Joe-Clayton copula and the Gumbel copula have the same RoV but the VaR for Joe-Clayton
copula is more robust than the Gumbel copula. This phenomenon can be seen more clearly in
both the periods of mid 2010 and late 2011, when the volatility occurs in portfolio
returns.

\section{Concluding remarks}
\label{sec:conclusion}

A general approach for covariate-dependent copula forecasting is proposed. The copula
features, as well as the features in the margins are linked to covariates. We use an
efficient Bayesian MCMC method to sample the posterior distribution and simultaneously
perform variable selection in all components of the model. Both simulation studies and
application to the daily returns of the S\&P 100 and the S\&P 600 indices show the
advantages of this approach in terms of understanding the bivariate dependence via known
covariates, improving the performances in both tail-dependence forecasting and predictive
density forecasting, and evaluating Value-at-Risk.  While bivariate applications of
copula-based models are still popular, high-dimensional covariate-dependent copula
modeling with discrete margins is also a direction for further research by using vine
copula constructions or data augmentation such as that exhibited in
\citet{pitt2006efficient}, \citet{smith2012estimation}, as well as
\citet{panagiotelis2017model}, but requires further extensive studies.

\section*{Acknowledgements}

The authors are grateful to Handling Editor Professor Dick van Dijk, Associate Editor, the
anonymous reviewer and Professor Mattias Villani for their helpful comments that improve
the contents of this paper. Feng Li and Yanfei Kang's research were supported by the
National Natural Science Foundation of China (No.~$11501587$ and No.~$11701022$,
respectively). The authors appreciate the support of this work from the ``Fundamental
Research Funds for the Central Universities'' (YWF-17-XGB-005).

\section*{References}
\bibliographystyle{elsarticle-harv}
\biboptions{authoryear}
\bibliography{full,References}

\begin{thebibliography}{47}
\expandafter\ifx\csname natexlab\endcsname\relax\def\natexlab#1{#1}\fi
\expandafter\ifx\csname url\endcsname\relax
  \def\url#1{\texttt{#1}}\fi
\expandafter\ifx\csname urlprefix\endcsname\relax\def\urlprefix{URL }\fi

\bibitem[{Aas et~al.(2009)Aas, Czado, Frigessi, and Bakken}]{aas2009pair}
Aas, K., Czado, C., Frigessi, A., Bakken, H., Apr. 2009. {Pair-copula
  constructions of multiple dependence}. Insurance: Mathematics and Economics
  44~(2), 182--198.

\bibitem[{Bollerslev(1986)}]{bollerslev1986generalized}
Bollerslev, T., 1986. {Generalized autoregressive conditional
  heteroskedasticity}. Journal of Econometrics 31~(3), 307--327.

\bibitem[{Clayton(1978)}]{clayton1978model}
Clayton, D., 1978. A model for association in bivariate life tables and its
  application in epidemiological studies of familial tendency in chronic
  disease incidence. Biometrika 65~(1), 141--151.

\bibitem[{Czado(2010)}]{czado2010pair}
Czado, C., 2010. Pair-copula constructions of multivariate copulas. In:
  Jaworski, P., Durante, F., H{\"a}rdle, W.~K., Rychlik, T. (Eds.), Copula
  theory and its applications: proceedings of the workshop held in Warsaw,
  25-26 September 2009. Springer, pp. 93--109.

\bibitem[{Czado et~al.(2012)Czado, Schepsmeier, and Min}]{czado2012maximum}
Czado, C., Schepsmeier, U., Min, A., 2012. Maximum likelihood estimation of
  mixed c-vines with application to exchange rates. Statistical Modelling
  12~(3), 229--255.

\bibitem[{Dobri{\'c} and Schmid(2005)}]{dobric2005nonparametric}
Dobri{\'c}, J., Schmid, F., 2005. Nonparametric estimation of the lower tail
  dependence $\lambda_{L}$ in bivariate copulas. Journal of Applied Statistics
  32~(4), 387--407.

\bibitem[{Draisma et~al.(2004)Draisma, Drees, Ferreira, and
  De~Haan}]{draisma2004bivariate}
Draisma, G., Drees, H., Ferreira, A., De~Haan, L., 2004. Bivariate tail
  estimation: dependence in asymptotic independence. Bernoulli 10~(2),
  251--280.

\bibitem[{Durrleman et~al.(2000)Durrleman, Nikeghbali, and
  Roncalli}]{durrleman2000simple}
Durrleman, V., Nikeghbali, A., Roncalli, T., 2000. A simple tranformation of
  copulas. Working paperGroupe de Recherce Operationelle, Credit Lyonnais,
  France.

\bibitem[{Embrechts et~al.(1997)Embrechts, Kl{\"u}ppelberg, and
  Mikosch}]{embrechts1997modelling}
Embrechts, P., Kl{\"u}ppelberg, C., Mikosch, T., 1997. Modelling extremal
  events. Vol.~33. Springer Science \& Business Media.

\bibitem[{Engle(2002)}]{engle2002dynamic}
Engle, R., 2002. Dynamic conditional correlation: A simple class of
  multivariate generalized autoregressive conditional heteroskedasticity
  models. Journal of Business \& Economic Statistics 20~(3), 339--350.

\bibitem[{Garc\'{\i}a et~al.(2013)Garc\'{\i}a, Gonz\'{a}lez-L\'{o}pez, and
  Nelsen}]{garcia2013new}
Garc\'{\i}a, J.~E., Gonz\'{a}lez-L\'{o}pez, V., Nelsen, R., Mar. 2013. {A new
  index to measure positive dependence in trivariate distributions}. Journal of
  Multivariate Analysis 115, 481--495.

\bibitem[{Geweke(2001)}]{geweke2001bayesian}
Geweke, J., 2001. Bayesian econometrics and forecasting. Journal of
  Econometrics 100~(1), 11--15.

\bibitem[{Geweke and Amisano(2010)}]{geweke2010comparing}
Geweke, J., Amisano, G., 2010. Comparing and evaluating {B}ayesian predictive
  distributions of asset returns. International Journal of Forecasting 26~(2),
  216--230.

\bibitem[{Geweke and Keane(2007)}]{geweke2007smoothly}
Geweke, J., Keane, M., 2007. {Smoothly mixing regressions}. Journal of
  Econometrics 138~(1), 252--290.

\bibitem[{Hilal et~al.(2011)Hilal, Poon, and Tawn}]{hilal2011hedging}
Hilal, S., Poon, S.-H., Tawn, J., Sep. 2011. {Hedging the black swan:
  Conditional heteroskedasticity and tail dependence in S\&P500 and VIX}.
  Journal of Banking \& Finance 35~(9), 2374--2387.

\bibitem[{Huang et~al.(2009)Huang, Lee, Liang, and Lin}]{huang2009estimating}
Huang, J.-J., Lee, K.-J., Liang, H., Lin, W.-F., 2009. Estimating value at risk
  of portfolio by conditional copula-{GARCH} method. Insurance: Mathematics and
  economics 45~(3), 315--324.

\bibitem[{Jaworski et~al.(2010)Jaworski, Durante, H{\"a}rdle, and
  Rychlik}]{jaworski2010copula}
Jaworski, P., Durante, F., H{\"a}rdle, W.~K., Rychlik, T., 2010. Copula theory
  and its applications: proceedings of the workshop held in Warsaw, 25-26
  September 2009. Vol. 198. Springer.

\bibitem[{Joe(1993)}]{joe1993parametric}
Joe, H., 1993. Parametric families of multivariate distributions with given
  margins. {Journal of Multivariate Analysis} 46~(2), 262--282.

\bibitem[{Joe(1997)}]{joe1997multivariate}
Joe, H., 1997. Multivariate models and dependence concepts. Chapman \& Hall,
  London.

\bibitem[{Joe(2005)}]{joe2005asymptotic}
Joe, H., Jun. 2005. {Asymptotic efficiency of the two-stage estimation method
  for copula-based models}. Journal of Multivariate Analysis 94~(2), 401--419.

\bibitem[{Kass(1993)}]{kass1993bayes}
Kass, R., 1993. {Bayes factors in practice}. Journal of the Royal Statistical
  Society. Series D (The Statistician) 42~(5), 551--560.

\bibitem[{Kastner and Fr{\"u}hwirth-Schnatter(2014)}]{kastner2014ancillarity}
Kastner, G., Fr{\"u}hwirth-Schnatter, S., 2014. Ancillarity-sufficiency
  interweaving strategy (asis) for boosting mcmc estimation of stochastic
  volatility models. Computational Statistics \& Data Analysis 76, 408--423.

\bibitem[{Kurowicka and Joe(2010)}]{kurowicka2010dependence}
Kurowicka, D., Joe, H., 2010. Dependence modeling: vine copula handbook. World
  Scientific.

\bibitem[{Ledford and Tawn(1997)}]{ledford1997modelling}
Ledford, A.~W., Tawn, J.~a., May 1997. {Modelling Dependence within Joint Tail
  Regions}. Journal of the Royal Statistical Society: Series B (Statistical
  Methodology) 59~(2), 475--499.

\bibitem[{Li et~al.(2010)Li, Villani, and Kohn}]{li2010flexible}
Li, F., Villani, M., Kohn, R., 2010. Flexible modeling of conditional
  distributions using smooth mixtures of asymmetric student t densities.
  Journal of Statistical Planning and Inference 140~(12), 3638--3654.

\bibitem[{Lucas et~al.(2014)Lucas, Schwaab, and Zhang}]{lucas2014conditional}
Lucas, A., Schwaab, B., Zhang, X., 2014. Conditional euro area sovereign
  default risk. Journal of Business \& Economic Statistics 32~(2), 271--284.

\bibitem[{Mardia and Kent(1979)}]{mardia1979multivariate}
Mardia, K., Kent, J., 1979. {Multivariate analysis}. Academic Press, London.

\bibitem[{Melino and Turnbull(1990)}]{melino1990pricing}
Melino, A., Turnbull, S.~M., 1990. Pricing foreign currency options with
  stochastic volatility. Journal of Econometrics 45~(1), 239--265.

\bibitem[{Nelsen(2006)}]{nelsen2006introduction}
Nelsen, R., 2006. {An introduction to copulas}. Springer Verlag.

\bibitem[{Nott and Kohn(2005)}]{nott2005adaptive}
Nott, D., Kohn, R., 2005. Adaptive sampling for {Bayes}ian variable selection.
  Biometrika 92~(4), 747--763.

\bibitem[{Panagiotelis et~al.(2017)Panagiotelis, Czado, Joe, and
  St{\"o}ber}]{panagiotelis2017model}
Panagiotelis, A., Czado, C., Joe, H., St{\"o}ber, J., 2017. Model selection for
  discrete regular vine copulas. Computational Statistics \& Data Analysis 106,
  138--152.

\bibitem[{Patton(2006)}]{patton2006modelling}
Patton, A., 2006. Modelling asymmetric exchange rate dependence. International
  economic review 47~(2), 527--556.

\bibitem[{Patton(2012)}]{patton2012copula}
Patton, A., 2012. Copula methods for forecasting multivariate time series. In:
  Handbook of economic forecasting. Vol.~2. Elsevier Oxford, pp. 899--960.

\bibitem[{Pitt et~al.(2006)Pitt, Chan, and Kohn}]{pitt2006efficient}
Pitt, M., Chan, D., Kohn, R., Sep. 2006. {Efficient {Bayes}ian inference for
  {Gauss}ian copula regression models}. Biometrika 93~(3), 537--554.

\bibitem[{Poon et~al.(2003)Poon, Rockinger, and Tawn}]{poon2003extreme}
Poon, S.-H., Rockinger, M., Tawn, J., 2003. Extreme value dependence in
  financial markets: Diagnostics, models, and financial implications. The
  Review of Financial Studies 17~(2), 581--610.

\bibitem[{Richardson and Green(1997)}]{richardson1997bayesian}
Richardson, S., Green, P., 1997. {On {Bayes}ian analysis of mixtures with an
  unknown number of components (with discussion)}. Journal of the Royal
  Statistical Society. Series B. Statistical Methodology 59~(4), 731--792.

\bibitem[{Schmidt and Stadtmuller(2006)}]{schmidt2006non}
Schmidt, R., Stadtmuller, U., Jun. 2006. {Non-parametric Estimation of Tail
  Dependence}. Scandinavian Journal of Statistics 33~(2), 307--335.

\bibitem[{Scott et~al.(2010)Scott, Berger, et~al.}]{scott2010bayes}
Scott, J.~G., Berger, J.~O., et~al., 2010. Bayes and empirical-bayes
  multiplicity adjustment in the variable-selection problem. The Annals of
  Statistics 38~(5), 2587--2619.

\bibitem[{Siburg et~al.(2015)Siburg, Stoimenov, and
  Wei{\ss}}]{siburg2015forecasting}
Siburg, K.~F., Stoimenov, P., Wei{\ss}, G.~N., 2015. Forecasting
  portfolio-value-at-risk with nonparametric lower tail dependence estimates.
  Journal of Banking \& Finance 54, 129--140.

\bibitem[{Sklar(1959)}]{sklar1959fonctions}
Sklar, A., 1959. Fonctions de répartition à n dimensions et leurs marges.
  Publications de l'Institut de Statistique de L'Université de Paris 8,
  229--231.

\bibitem[{Smith and Khaled(2012)}]{smith2012estimation}
Smith, M., Khaled, M., 2012. Estimation of copula models with discrete margins
  via {B}ayesian data augmentation. Journal of the American Statistical
  Association 107~(497), 290--303.

\bibitem[{Smith(2015)}]{smith2015copula}
Smith, M.~S., 2015. Copula modelling of dependence in multivariate time series.
  International Journal of Forecasting 31~(3), 815--833.

\bibitem[{Tran et~al.(2014)Tran, Giordani, Mun, Kohn, and
  Pitt}]{tran2014copula}
Tran, M.-N., Giordani, P., Mun, X., Kohn, R., Pitt, M.~K., 2014. Copula-type
  estimators for flexible multivariate density modeling using mixtures. Journal
  of Computational and Graphical Statistics 23~(4), 1163--1178.

\bibitem[{Villani et~al.(2009)Villani, Kohn, and
  Giordani}]{villani2009regression}
Villani, M., Kohn, R., Giordani, P., Dec. 2009. {Regression density estimation
  using smooth adaptive {Gauss}ian mixtures}. Journal of Econometrics 153~(2),
  155--173.

\bibitem[{Villani et~al.(2012)Villani, Kohn, and Nott}]{villani2012generalized}
Villani, M., Kohn, R., Nott, D.~J., 2012. Generalized smooth finite mixtures.
  Journal of Econometrics 171~(2), 121--133.

\bibitem[{Yau et~al.(2003)Yau, Kohn, and Wood}]{yau2003bayesian}
Yau, P., Kohn, R., Wood, S., 2003. Bayesian variable selection and model
  averaging in high-dimensional multinomial nonparametric regression. Journal
  of Computational and Graphical Statistics 12~(1), 23--54.

\bibitem[{Yu and Meyer(2006)}]{yu2006multivariate}
Yu, J., Meyer, R., 2006. Multivariate stochastic volatility models: {B}ayesian
  estimation and model comparison. Econometric Reviews 25~(2-3), 361--384.

\end{thebibliography}

\appendix

\section{New properties of Joe-Clayton copula}
\label{sec:bb7-prope}

The Joe-Clayton copula has some unique attributes. The upper tail-dependence and lower
tail-dependence are not functionally dependent. The Clayton copula
\citep{clayton1978model} and the B5 copula \citep{joe1993parametric} are special cases of
the Joe-Clayton copula. All of them belong to a more general class of Archimedean
copulas. Furthermore, we also find the following new properties.

\begin{enumerate}[noitemsep]
\item The inequality holds $0\leq\lambda_{L}\leq2^{1/2-1/(2\tau)}$ when the lower
  tail-dependence is not extremely high. We say that $\lambda_{L}$ and $\tau$ are
  \emph{variationally dependent}. The proof is non-trivial, but we have verified the
  inequality numerically in a very careful way. We discover the bound of the inequality
  through the limit of $\tau(\lambda_{L},\lambda_{U})$ when $\lambda_{U}\to0$. See
  Figure~\ref{fig:tau} for illustration.

\item When $\lambda_{L}\to0$ (i.e. $\delta\to0$), we have
  \begin{align*}
    \tau\to1-\frac{2H(2/\theta)-2}{2-\theta}=1-\frac{2H(2\log(2-\lambda_{u})/\log2)
    -2}{2-\log2/\log(2-\lambda_{u})},
  \end{align*}
  and
  \begin{align*}
    \frac{\partial\tau}{\partial\theta}\to\frac{2(1-H(2/\theta))}{(\theta-2)^{2}}
    -\frac{4\psi_{1}(2/\theta+1)}{(\theta-2)\theta^{2}},
  \end{align*}
  where $H(\cdot)$ is the harmonic number. When $\theta\to2$
  (i.e. $\lambda_{U}\to2-\sqrt{2}\approx0.59)$, we have $\tau\to2-\pi^{2}/6\approx0.36$
  and $\partial\tau/\partial\theta\to\pi^{2}/12-Zeta(3)/2\approx0.22$, where $Zeta(\cdot)$
  is the Riemann zeta function. This property allows us to have a more numerically robust
  posterior inference when a nearly non-dependent situation applies.

\item We also derive the analytical gradients for the copula density with respect to the
  Kendall's $\tau$ and tail-dependence of Joe-Clayton copula in \ref{sec:grad-bb7}, which
  are used to construct efficient proposal distributions for the MCMC.

\end{enumerate}

The Joe-Clayton copula can only determine positive correlations. If the relationship
between two variables is negative, we simply need to rotate the axes of the copula and the
estimation procedure remains the same. For example, for copulas rotated by 90 degrees, $u$
has to be set to $1-u$. For 270 degrees, let $v$ be $1-v$.  For 180 degrees, set $u$ and
$v$ to $1-u$ and $1-v$, respectively. See \citet{durrleman2000simple} for the proof that
after this transformation, it is still a copula and for other possible transformations to
extend the current bivariate copula with more desired properties. In the financial
application exhibited in Section \ref{sec:application}, the correlations are known to be
positive, but modeling negative correlations requires no extra work with our developed R
package.

\section{The MCMC details}

In this section, we briefly present the MCMC details. The MCMC implementation is
straightforward, but requires great care of the proposal distribution in the
Metropolis--Hastings algorithm.

\subsection{The chain rule for the features in copula models}
\label{sec:chain-rule}

We use the finite-step Newton method embedded in the Metropolis-Hastings algorithm that
requires the analytical gradient for the posterior with respect to the features of
interest in the marginal and copula components. The chain rule of gradient conveniently
modularizes the copula model and reduces the complexity of the gradient calculation.
\begin{align*}
  \frac{\partial\log c(u_{1},...,u_{M},\lambda_{L},\tau)}{\partial\lambda_{L}}= &
                                                                                  \frac{\partial\log
                                                                                  c(u_{1},...,u_{M},\theta,\delta)}{\partial\delta}\times\left(\frac
                                                                                  {\partial\lambda_{L}}{\partial\delta}\right)^{-1},\\
  \frac{\partial\log c(u_{1},...,u_{m},\lambda_{L},\tau)}{\partial\tau}= &
                                                                           \frac{\partial\log
                                                                           c(u_{1},...,u_{m},\theta,\delta)}{\partial\theta}\times\left(\frac{\partial\tau
                                                                           (\theta,\delta)}{\partial\theta}\right)^{-1},\\
  \frac{\partial\log c(u_{1},...,u_{M},\lambda_{L},\tau)}{\partial\varphi_{m}}= &
                                                                                  \frac{\partial\log c(u_{1},...,u_{M},\theta,\delta)}{\partial
                                                                                  u_{m}}\times\frac{\partial u_{m}}{\partial\varphi_m} + \frac{\partial \log
                                                                                  p_m(y_m,\varphi_m)}{\partial \varphi_m},
\end{align*}
where $\varphi_{m}$ is any feature in the $m$:th marginal density $p_m(y_m,\varphi_m)$ and
$u_{m}$ is the CDF function of its marginal density. The parameters $\theta$ and $\delta$
are the intermediate parameters that link the dependence and correlations with the
traditional parametrization for copulas. Particularly, obtaining the gradient for the
Student's \emph{t}-copula requires the following extra decomposition,
\begin{align*}
  \frac{\partial \log c_t(u_{1},...,u_{M},\lambda_L,\tau)}{\partial u_m} =
  \frac{\partial \log
  p_t(x_1,...,x_M,\nu)}{\partial x_m}/\frac{\partial u_m}{\partial x_m},
\end{align*}
where $u_m = \int p_m(x_m,\nu) dx_m$ are the CDF for univariate Student $t$ density with
$\nu$ degrees of freedom in the $m$th margin. Gradients for other elliptical copulas are
obtained in a similar way.

The MCMC algorithm requires evaluating $\tau_{\lambda_{L}}^{-1}$ excessively.  It can be
evaluated numerically or through a dictionary-lookup method. In practice, we have found
that the dictionary-lookup method is particularly fast and robust. Modeling the upper
tail-dependence can be done in the same manner.

Our model in Section \ref{sec:copula-model} is covariate-dependent.  Let
$l(\varphi)=x'\beta$ be the link function where $\varphi$ is the feature of interest. The
gradient expression can be written as
\begin{align*}
  \frac{\partial\log c(u_{1},...,u_{M},\varphi)}{\partial\beta}=\frac{\partial\log
  c(u_{1},...,u_{M},\varphi)}{\partial\varphi}\times\left(\frac{\partial
  l(\varphi)}{\partial\varphi}\right)^{-1}\times\frac{\partial
  x'\beta}{\partial\beta}.
\end{align*}

When the conditional link function is used, e.g. $\tau$ depends on $\lambda_{L}$ in our
model in the link function $l(\tau|\lambda_{L})=x'\beta$, the gradient for $\lambda_{L}$
is slightly more complicated. One needs to write $\tau$ as a function of $\lambda_{L}$
with the link function and substitute it into the copula density. The gradient for
$\lambda_{L}$ is obtained thereafter. The details are straightforward, but lengthy, and
are omitted here.

\subsection{Gradients for features in the Joe-Clayton copula}
\label{sec:grad-bb7}

The gradient for the Joe-Clayton copula w.r.t lower tail-dependence $\lambda_{L}$ can be
decomposed as
\begin{align*}
  \frac{\partial\log c(u,v,\theta,\delta)}{\partial\delta}= & -\log T_{1}(u)-\log
                                                              T_{1}(v)-\frac{2(1+\delta)\Delta_{1}}{\delta
                                                              L_{1}}-(\frac{1}{\theta}-2)\frac{\log
                                                              L_{1}-\delta\Delta_{1}/L_{1}}{\delta^{2}(L_{1}^{1/\delta}-1)}\\
                                                            & +\frac{2\log
                                                              L_{1}}{\delta^{2}}+\frac{L_{1}^{1/\delta}-(1+\delta)L_{1}^{1/\delta}(\log
                                                              L_{1}-\delta\Delta_{1}/L_{1})/\delta^{2}-1}{(1+\delta)L_{1}^{1/\delta}-\delta-1
                                                              /\theta},
\end{align*}
where
$\Delta_{1}=\partial L_{1}/\partial\delta=-T_{1}(u)^{-\delta}\log
T_{1}(u)-T_{1}(v)^{-\delta}\log T_{1}(v)$.  Furthermore,
$\partial\lambda_{L}/\partial\delta=2^{-1/\delta}\log(2)/\delta^{2}.$

The gradient for Joe-Clayton copula with respect to the Kendall's $\tau$ is decomposed as
\begin{align*}
  \frac{\partial\log c(u,v,\theta,\delta)}{\partial\theta}= &
                                                              -(1+\delta)\Delta_{2}(0)+\Delta_{3}(u)+\Delta_{3}(v)+2(1+\delta)\Delta_{2}(
                                                              -\delta)/L_{1}\\
                                                            &
                                                              +\frac{(1-2\theta)\Delta_{2}(1/\delta)}{(1-L_{1}^{-1/\delta})\theta\delta}
                                                              -\frac{\log(1-L_{1}^{-1/\delta})}{\theta^{2}}\\
                                                            & +\frac{(1+\delta)L_{1}^{1/\delta}+\theta(1+\delta)/\delta
                                                              L_{1}^{1/\delta-1}\Delta_{2}(-1/\delta)-\delta}{(1+\delta)\theta
                                                              L_{1}^{1/\delta}-\theta\delta-1},
\end{align*}
where
$\Delta_{2}(d)=-T_{1}(u)^{d-1}(1-u)^{\theta}\log(1-u)-T_{1}(v)^{d-1}(1-v)^
{\theta}\log(1-v)$ and
$\Delta_{3}(s)=\partial\log T_{2}(s)/\partial\theta=(1-s)^{\theta-1}\log(1-s)/T_{2}(s)$.
Furthermore,
\begin{align*}
  \frac{\partial\tau(\theta,\delta)}{\partial\theta} = \begin{cases}
    -2/[(\theta-2)^{2}\delta]-8B(2+\delta,2/\theta-1)\left[\theta+\psi(2/\theta-1)
      -\psi(2/\theta+\delta+1)\right]/(\theta^{4}\delta),\\
    \hspace{0.7\textwidth}1\leq\theta<2;\\
    -[12+24\psi(1)+6\psi^{2}(1)+\pi^{2}-12(2+\psi(1))\psi(2+\delta)+6\psi^{2}(2
    +\delta)-6\psi_{1}(2+\delta)]/(24\delta),\\
    \hspace{0.7\textwidth}\theta=2;\\
    \frac{\left({
          \begin{array}{c}
            -2(2+\delta)\theta^{4}B(1+\delta+2/\theta,2-2/\theta)-8\pi^{2}(\theta-2)^{2
            }\cos(2\pi/\theta)/\sin^{2}(2\pi/\theta)\\
            -8\pi(\theta-2)^{2}[\psi(1+\delta+2/\theta)-\psi(2-2/\theta)-\theta]/\sin(2\pi
            /\theta)
          \end{array}}\right)}
    {\begin{array}{l}
       \delta(2+\delta)(\theta-2)^{2}\theta^{4}B(1+\delta+2/\theta,2-2/\theta)
       \end {array}},\\
     \hspace{0.7\textwidth}\theta>2,
   \end{cases}
\end{align*}
where $\psi_{1}(\cdot)$ is the trigamma function. then gradient for the case $\theta=2$
can be obtained by taking the limiting result from the cases of $1\leq\theta<2$ or
$\theta>2$ when $\theta\to2$,

\begin{align*}
  \frac{\partial\tau(\theta,\delta)}{\partial\delta}=\begin{cases}
    -2/[(\theta-2)\delta^{2}]+4B(2+\delta,2/\theta-1)[\psi(2+\delta)-\psi(2/\theta
    +\delta+1)-1/\delta]/(\theta^{2}\delta),\\
    \hspace{0.7\textwidth}1\leq\theta<2;\\
    {}[\psi(2+\delta)-\delta\psi_{1}(2+\delta)-\psi(1)-1]/\delta^{2},{\hspace{5.2cm
      }}\theta=2;\\
    \begin{aligned}-\frac{2}{(\theta-2)\delta^{2}}-\frac{4\pi[\psi(3+\delta)-\psi(2
        /\theta+\delta+1)-2(1+\delta)/(2\delta+\delta^{2})]}{(2+\delta)\delta\theta^{2
        }\sin(2\pi/\theta)B(1+\delta+2/\theta,2-2/\theta)},\end{aligned}
    \\
    \hspace{0.7\textwidth}\theta>2.
  \end{cases}
\end{align*}

For the Joe-Clayton copula, $u$ and $v$ are exchangeable, and we only present the
derivative with respect to $u$:

\begin{align*}
  \frac{\partial\log c(u,v,\theta,\delta)}{\partial u}= &
                                                          (1+\delta)\theta\Delta_{4}(0)+(1-\theta)[(1-u)^{\theta-2}/T_{2}(u)+(1-v)^
                                                          {\theta-2}/T_{2}(v)]\\
                                                        &
                                                          -2(1+\delta)\Delta_{4}(-\delta)/L_{1}-(1/\theta-2)L_{1}^{-1/\delta-1}\Delta_{4}
                                                          (-\delta)/(1-L_{1}^{-1/\delta})\\
                                                        & -(1+\delta)\theta
                                                          L_{1}^{1/\delta-1}\Delta_{4}(-\delta)/\left[(1+\delta)\theta
                                                          L_{1}^{1/\delta}-\theta\delta-1\right]
\end{align*}
where
$\Delta_{4}(d)=-T_{1}(u)^{d-1}(1-u)^{\theta-1}\theta-T_{1}(v)^{d-1}(1-v)^
{\theta-1}\theta$.

\subsection{Gradients for features in Student's \emph{t}-copula and the
  covariate-dependent structure}
\label{sec:grad-mvt}

The gradient for the Student's \emph{t}-copula w.r.t. the lower tail dependence
$\lambda_{Lij}$ for the $i$th and $j$th margins can be obtained via the following chain
rule,
\begin{align*}
  \frac{\partial \log c_t(\bm{u},\lambda_{Lij},\tau)}{\partial \lambda_{Lij}} =
  \frac{\partial \log
  p_t(\bm{x},\nu,\rho_{ij})}{\partial \nu}/\frac{\partial \lambda_{Lij}}{\partial
  \nu},
\end{align*}
where the tail-dependence for $i$th and $j$th margins ($\lambda_{Lij}$) are
\citep{embrechts1997modelling}
\begin{align*}
  \lambda_{Lij} = \frac{\int _{\pi/4 - \mathrm{arcsin}
  (\rho_{ij})/2}^{\pi/2}\mathrm{cos}^{\nu}(t)dt}{\int _0^{\pi} \mathrm{cos}^{\nu}
  (t) d t},
\end{align*}
where $\rho_{ij}$ is the correlation coefficient for $i$th and $j$th margins.

In a $M$-dimensional Student's \emph{t}-copula, the covariate-dependent structure can be
conveniently represented in a matrix form as
\begin{align*}
  \mathrm{vec}(\bm{T}) = l^{-1}([\bm{I}\otimes \bm{X}]\mathrm{vec}\bm{B}),
\end{align*}
where $\bm{T}$ is a $n\times [M(M-1)/2]$ dependence matrix, $\bm{I}$ is a $M(M-1)/2$
dimensional identity matrix, $\bm{X}$ is the set of covariates used in all marginal
models, and $\bm{B}$ is the corresponding coefficient matrix. Other copula features can be
linked to covariates in the same way. This representation allows us to jointly estimate
the correlation and tail-dependences. Note that there is a curse of dimensionality in
higher dimensional modeling with covariate-dependent structures. Currently, limited tests
for a smaller number of dimensions (less than $10$) are tried with the Student's
$t$-copula. We find that the computation time dramatically increases as the number of
dimension grows.

\subsection{Gradients for features in the marginal distributions}
\label{sec:grad-margi}

The direct derivatives of CDF function and PDF functions with respect to their features
are straightforward for most densities.  We only document the split-\emph{t} case for CDF
functions.  Let $I=\kappa$ if $y>\mu$ and $I=1$ elsewhere, $J=1$, if $y>\mu$ and $J=-1$
elsewhere, and $A=I^{2}\nu\phi^{2}/[(y-\mu)^{2}+I^{2}\nu\phi^{2}]$. The gradient for the
split-\emph{t }CDF function with respect to its features $\mu,\phi,\kappa,\nu$ are as
follows,

\begin{align*}
  \frac{\partial
  u_{split-t}(y,\mu,\phi,\kappa,\nu)}{\partial\mu}=&-\frac{2I\sqrt{\frac{1}{(y
                                                     -\mu)^{2}+I^{2}\nu\phi^{2}}}A^{\nu/2}}{(1+\kappa)\text{Beta}\left[\frac{\nu}{2}
                                                     ,\frac{1}{2}\right]},\\
  \frac{\partial
  u_{split-t}(y,\mu,\phi,\kappa,\nu)}{\partial\phi}=&-\frac{2I(y-\mu)\sqrt{\frac
                                                      {1}{(y-\mu)^{2}+I^{2}\nu\phi^{2}}}A^{\nu/2}}{(1+\kappa)\phi\text{Beta}\left
                                                      [\frac{\nu}{2},\frac{1}{2}\right]},\\
  \frac{\partial u_{split-t}(y,\mu,\phi,\kappa,\nu)}{\partial\phi}=&\begin{cases}
    -\frac{\text{Beta}_{R}\left[A,\frac{\nu}{2},\frac{1}{2}\right]}{(1+\kappa)^{2}}
    & y<\mu\\
    -\frac{2(1+\kappa)(y-\mu)\sqrt{\frac{1}{(y-\mu)^{2}+\kappa^{2}\nu\phi^{2}}}A^
      {\nu/2}+\text{Beta}\left[A,\frac{\nu}{2},\frac{1}{2}\right]}{(1+\kappa)^{2
      }\text{Beta}\left[\frac{\nu}{2},\frac{1}{2}\right]} & \text{elsewhere,}
  \end{cases}\\
  \frac{\partial u_{split-t}(y,\mu,\phi,\kappa,\nu)}{\partial\phi}= &
                                                                      \frac{I}{2(1+\kappa)\nu^{2}\text{Beta}\left[\frac{\nu}{2},\frac{1}{2}\right]
                                                                      }\left\{
                                                                      4JA^{\nu/2}{}_{p}F_{q}\left[\left\{
                                                                      \frac{1}{2},\frac{\nu}{2},\frac{\nu}{2}\right\}
                                                                      ,\left\{
                                                                      1+\frac{\nu}{2},1+\frac{\nu}{2}\right\} ,A\right]\right.\\
                                                   &
                                                     \left.+\nu\left(-2(y-\mu)\sqrt{\frac{1}{(y-\mu)^{2}+\kappa^{2}\nu\phi^{2}}}A^
                                                     {\nu/2}\right.\right.\\
                                                   & -\nu J\text{
                                                     Beta}\left[A,\frac{\nu}{2},\frac{1}{2}\right]\left(\log\left(A\right)-\text
                                                     {\ensuremath{\psi}}\left(\frac{\nu}{2}\right)+\text{\ensuremath{\psi}}\left
                                                     (\frac{1+\nu}{2}\right)\right)\left.\right.\Bigg\},
\end{align*}
where $\mathrm{Beta}_{R}$ is the regularized beta function, $_{p}F_{q}$ is the generalized
hypergeometric function.

However, there are exceptions when this derivative is numerically unstable in practice. In
this situation, we propose an alternative approach.  Note that
\begin{align*}
  \frac{\partial u}{\partial\varphi}=\frac{\partial
  F(y,\varphi)}{\partial\varphi}=\int_{-\infty}^{y}\frac{\partial
  f(x,\varphi)}{\partial\varphi}\mathrm{d}x\label{eq:derive},
\end{align*}
where $u=F(y,\varphi)$ is the CDF function of density $f(y,\varphi)$, and calculating
$\partial f(y,\varphi)/\partial\varphi$ is usually easier than
$\partial F(y,\varphi)/\partial\varphi$. When the integral cannot be easily obtained
analytically, numerical methods can be applied in the last stage.

\end{document}